\documentclass[graybox]{svmult}

\usepackage[T1]{fontenc}
\usepackage{type1cm}        % activate if the above 3 fonts are
\usepackage{makeidx}         % allows index generation
\usepackage{graphicx}        % standard LaTeX graphics tool
\usepackage{multicol}        % used for the two-column index
\usepackage[bottom]{footmisc}% places footnotes at page bottom

\usepackage{newtxtext}       % 
\usepackage[varvw]{newtxmath}       % selects Times Roman as basic font

\usepackage{booktabs}
\usepackage{url}

\usepackage{csquotes}
\graphicspath{{../../graphics}}
\usepackage{hyperref}
 \hypersetup{
     colorlinks=true,
     linkcolor=black,
     filecolor=black,
     citecolor = black,      
     urlcolor=black
     }
\usepackage{siunitx}
\usepackage{subcaption}	
    \def\equationautorefname~#1\null{Eq. (#1)\null}

\definecolor{tud1a}{HTML}{5D85C3}
\definecolor{tud2atrue}{HTML}{009CDA}%fnb blau, my bright blue
\definecolor{tud2a}{rgb}{0.00000,0.44700,0.74100}% % #0072BD
\definecolor{tud2aHEX}{HTML}{0072BD}
\definecolor{tud3a}{HTML}{50B695}
\definecolor{tud3ai}{HTML}{009a67} %my turkis
\definecolor{tud4a}{HTML}{AFCC50}
\definecolor{tud5a}{HTML}{DDDF48}
\definecolor{tud6a}{HTML}{FFE05C}
\definecolor{tud6b}{HTML}{FDCA00}%gelb
\definecolor{tud7b}{HTML}{F5A300} % my orange
\definecolor{tud8a}{HTML}{EE7A34}
\definecolor{tud8b}{HTML}{EC6500}%orange
\definecolor{tud9a}{HTML}{E9503E}
\definecolor{tud11b}{HTML}{721085}%dunkellila
\definecolor{tud1b}{HTML}{005AA9} % tiefblau
\definecolor{tud3b}{HTML}{009D81}
\definecolor{tud4b}{HTML}{99C000}
\definecolor{tud4c}{HTML}{7FAB16}% grÃŒn
\definecolor{tud4d}{HTML}{6A8B22}
\definecolor{tud9b}{HTML}{E6001A}%hellrot
\definecolor{tud9c}{HTML}{B90F22} % tiefrot
\definecolor{tud10a}{HTML}{A60084} % my red
\definecolor{sourcecode}{HTML}{E9E9E9}
\definecolor{mygrey}{HTML}{C0C0C0}

\definecolor{tud2a}{HTML}{0072BD}%
\definecolor{tud10a}{HTML}{BE1622}

\newcommand{\vect}[1]{\boldsymbol{#1}}
\newcommand{\Arch}{\operatorname{\mathit{A\kern-.06em r}}} % http://de.wikipedia.org/wiki/Archimedes-Zahl
\newcommand{\Biot}{\operatorname{\mathit{B\kern-.06em i}}} % http://de.wikipedia.org/wiki/Biot-Zahl
\newcommand{\Cauc}{\operatorname{\mathit{C\kern-.07em a}}} % http://de.wikipedia.org/wiki/Cauchy-Zahl
\newcommand{\Damk}{\operatorname{\mathit{D\kern-.06em a}}} % http://de.wikipedia.org/wiki/Damk%C3%B6hler-Zahl
\newcommand{\Eule}{\operatorname{\mathit{E\kern-.03em u}}} % http://de.wikipedia.org/wiki/Euler-Zahl
\newcommand{\Four}{\operatorname{\mathit{F\kern-.10em o}}} % http://de.wikipedia.org/wiki/Fourier-Zahl
\newcommand{\Frou}{\operatorname{\mathit{F\kern-.07em r}}} % http://de.wikipedia.org/wiki/Froude-Zahl
\newcommand{\Gras}{\operatorname{\mathit{G\kern-.05em r}}} % http://de.wikipedia.org/wiki/Grashof-Zahl
\newcommand{\Karl}{\operatorname{\mathit{K\kern-.11em a}}} % http://de.wikipedia.org/wiki/Karlovitz-Zahl
\newcommand{\Knud}{\operatorname{\mathit{K\kern-.11em n}}} % http://de.wikipedia.org/wiki/Knudsen-Zahl
\newcommand{\Lewi}{\operatorname{\mathit{L\kern-.05em e}}} % http://de.wikipedia.org/wiki/Lewis-Zahl
\newcommand{\Mach}{\operatorname{\mathit{M\kern-.10em a}}} % http://de.wikipedia.org/wiki/Mach-Zahl
\newcommand{\Nuss}{\operatorname{\mathit{N\kern-.09em u}}} % http://de.wikipedia.org/wiki/Nusselt-Zahl
\newcommand{\Pecl}{\operatorname{\mathit{P\kern-.08em e}}} % http://de.wikipedia.org/wiki/P%C3%A9clet-Zahl
\newcommand{\Pran}{\operatorname{\mathit{P\kern-.03em r}}} % http://de.wikipedia.org/wiki/Prandtl-Zahl
\newcommand{\Rayl}{\operatorname{\mathit{R\kern-.04em a}}} % http://de.wikipedia.org/wiki/Rayleigh-Zahl
\newcommand{\Reyn}{\operatorname{\mathit{R\kern-.04em e}}} % http://de.wikipedia.org/wiki/Reynolds-Zahl
\newcommand{\Rich}{\operatorname{\mathit{R\kern-.06em i}}} 
\newcommand{\Schm}{\operatorname{\mathit{S\kern-.07em c}}} % http://de.wikipedia.org/wiki/Schmidt-Zahl
\newcommand{\Sher}{\operatorname{\mathit{S\kern-.07em h}}} % http://de.wikipedia.org/wiki/Sherwood-Zahl
\newcommand{\Stro}{\operatorname{\mathit{S\kern-.07em r}}} % http://de.wikipedia.org/wiki/Strouhal-Zahl
\newcommand{\Webe}{\operatorname{\mathit{W\kern-.14em e}}} % http://de.wikipedia.org/wiki/Weber-Zahl

\usepackage[maxbibnames=20, 
    natbib=true, 
    style=ieee,
    isbn=false, 
    url=false,
    giveninits=true,
    sortcites]{biblatex}   % Literaturverzeichnis
\makeindex             % used for the subject index
\usepackage{pgfplots}
\usepackage{pgffor}
\usepackage{etoolbox}
\usepackage{listofitems} % for \readlist to create arrays
\pgfplotsset{width=10cm,compat=1.18}
\usepgfplotslibrary{external, statistics, groupplots, fillbetween, colormaps}

\definecolor{color11a}{RGB}{236, 188, 0}
\definecolor{color12a}{RGB}{50, 67, 121}
\definecolor{color13}{RGB}{222, 222, 222}
\definecolor{color21a}{RGB}{92, 134, 196}
\definecolor{color22a}{RGB}{249, 156, 0}
\definecolor{color23}{RGB}{222, 222, 222}
\definecolor{color31}{RGB}{222, 222, 222}
\definecolor{color32}{RGB}{222, 222, 222}
\definecolor{color33a}{RGB}{191, 60, 60}

\colorlet{color11}{color11a!90!color13}
\colorlet{color12}{color12a!90!color13}
\colorlet{color21}{color21a!90!color13}
\colorlet{color22}{color22a!90!color13}
\colorlet{color33}{color33a!90!color13}

\definecolor{CPSgreen}{RGB}{22,164,138}
\definecolor{CPSlightblue}{RGB}{104,143,198}
\definecolor{CPSdarkblue}{RGB}{67,83,132}
\definecolor{CPSgrey}{RGB}{204, 204, 204}
\definecolor{CPSorange}{RGB}{246,163,21}
\definecolor{CPSred}{RGB}{194,76,76}

\usetikzlibrary{shapes}
\usetikzlibrary{arrows.meta} % for arrow size

\tikzset{>=latex} % for LaTeX arrow head
\tikzstyle{node}        =[thick, circle, draw=black, text = black, minimum size=20, inner sep=0.5, outer sep=0.6]
\tikzstyle{node icnn}   =[node, color=color22!10!black, fill=color22!25]
\tikzstyle{node inout} =[node, color=color21!10!black, fill=color21!25]
\tikzstyle{node layer}  =[node, rectangle, color=color33!10!black, fill=color33!25]
\tikzstyle{node encoder}  =[node, trapezium, rounded corners=3, color=CPSorange!10!black, fill=CPSorange!25, rotate=-90]
\tikzstyle{node decoder}  =[node encoder, rotate=-180]
\tikzstyle{node inout large}  =[node inout, rectangle, rounded corners=5, inner sep=5]

\tikzstyle{connect}=[thick, black, shorten <=1, shorten >=1, rounded corners=3] %,line cap=round
\tikzstyle{connect arrow}=[connect, ->, rounded corners=3]
\tikzstyle{connect arrow constraint}=[connect arrow, color22a, rounded corners=3]
\tikzstyle{connect constraint}=[connect, color22a, rounded corners=3] %,line cap=round

\usepackage{siunitx}
\usepackage{graphicx}

\usepackage{todonotes}

\DeclareMathOperator{\NRMSE}{NRMSE}
\DeclareMathOperator{\RMSE}{RMSE}
\DeclareMathOperator{\STD}{STD}

\DeclareMathOperator{\modela}{\mathcal{M}_{Q\mapsto T}}
\DeclareMathOperator{\modelb}{\mathcal{M}_{T\mapsto \sigma}}
\DeclareMathOperator{\generalmodel}{\mathcal{M}_{u\mapsto y}}
\begin{document}

\title*{Digital twin inference from multi-physical simulation data of DED additive manufacturing processes with neural ODEs}
\titlerunning{Digital twin inference from multi-physical simulation data of DED AM processes}
\author{Maximilian Kannapinn\orcidID{0000-0001-9342-0802},\\ Fabian Roth\orcidID{0009-0005-9492-6668} and\\ Oliver Weeger\orcidID{0000-0002-1771-8129}}
\institute{Maximilian Kannapinn \at Technical University of Darmstadt, Cyber-Physical Simulation, 
Dolivostraße 15, 64293 Darmstadt, \email{kannapinn@cps.tu-darmstadt.de}
\and  Fabian Roth \at Technical University of Darmstadt, Cyber-Physical Simulation, 
Dolivostraße 15, 64293 Darmstadt,  \email{roth@cps.tu-darmstadt.de}
\and  Oliver Weeger \at Technical University of Darmstadt, Cyber-Physical Simulation, 
Dolivostraße 15, 64293 Darmstadt,  \email{weeger@cps.tu-darmstadt.de}}
\maketitle

\abstract{A digital twin is a virtual representation that accurately replicates its physical counterpart, fostering bi-directional real-time data exchange throughout the entire process lifecycle. For Laser Directed Energy Deposition of Wire (DED-LB/w) additive manufacturing processes, digital twins may help to control the residual stress design in build parts. This study focuses on providing faster-than-real-time and highly accurate surrogate models for the formation of residual stresses by employing neural ordinary differential equations. The approach enables accurate prediction of temperatures and altered structural properties like stress tensor components. The developed surrogates can ultimately facilitate on-the-fly re-optimization of the ongoing manufacturing process to achieve desired structural outcomes. Consequently, this building block contributes significantly to realizing digital twins and the first-time-right paradigm in additive manufacturing.}

\section{Introduction}
\label{sec:1}

\begin{figure}[t]
    \centering
        \includegraphics[width=0.49\textwidth]{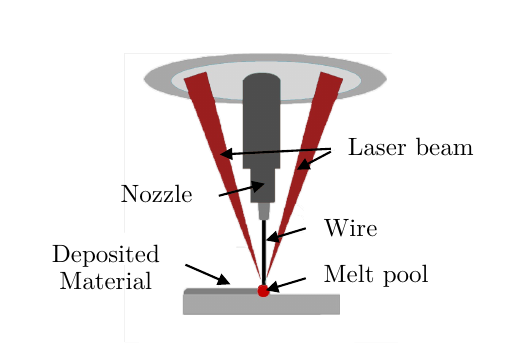}
        \includegraphics[width=0.49\textwidth]{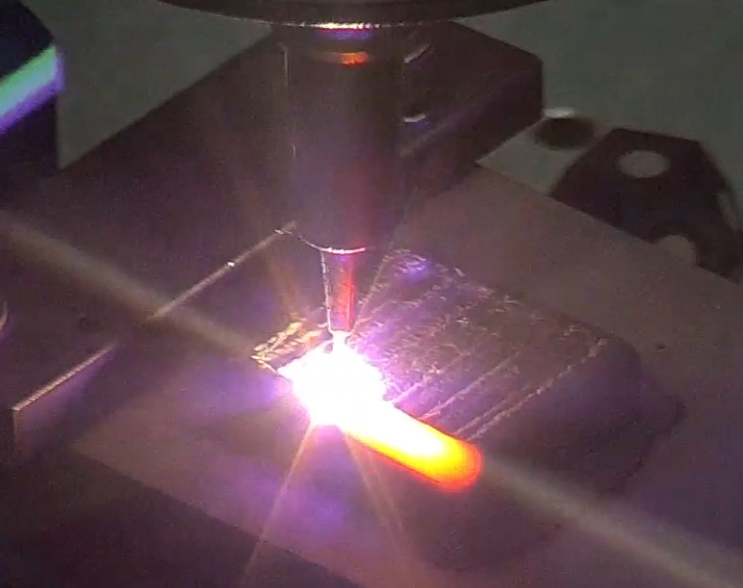}
        \caption{Illustration of the tool head of the DED-LB/w process and process in operation (kindly provided by PTW TEC at TU Darmstadt).}
    \label{fig:DEDprinciple}
\end{figure}

The technology of additive manufacturing (AM) of metals stands out due to its cost and environmentally efficient use of materials. In particular, the Laser Directed Energy Deposition of Wire (DED-LB/w, short just DED) process has a material efficiency of almost 100\% \cite{bernauer_segmentation_2024}. Figure~\ref{fig:DEDprinciple} illustrates the working principle in which a focused energy source is used to continuously melt a supplied metal wire, and deposition is performed in a directed manner on a substrate.

DED has considerable potential for applications in the aerospace, medical and energy sectors, as it enables the production of metal parts in almost any shape at high deposition rates. However, the material bond between the molten wire and the substrate is sensitive to disturbances and process instabilities. Intense thermal, mechanical and metallurgical interactions during the process lead to considerable residual stresses (RS) in the manufactured components. These have a major impact on the dimensional accuracy, corrosion resistance, fatigue strength against crack growth and general mechanical properties of the manufactured components. Consequently, the management of RS is crucial for optimizing the subsequent costs and improving the performance and quality of the products -- in short, they are one of the decisive factors as to whether DED will be used at scale~\cite{chen_review_2022, bastola_review_2023}.
Conventional approaches to eliminating RS are post-treatment strategies such as shot peening or stress relief annealing~\cite{chen_review_2022, bastola_review_2023}. However, all of these methods require more time and energy, which could prevent an already costly process from being widely adapted for industrial use~\cite{bartlett_overview_2019}.

This motivates the development of modern control methods based on digital twins to control the RS formation and functional design live during printing. 
The presented proof of concept study investigates the feasibility of neural ordinary differential equations (NODEs) surrogate models to accurately and swiftly predict the temperature and RS fields of a DED process.
To this aim, the remainder of this work is organized as follows:  As this work can be considered one of the first research endeavors for RS modeling through digital twins of the DED process with NODEs, this introduction comprises comprehensive information on the basic RS formation principles in \autoref{sec:1rs}, the state of the art of simulating RS in \autoref{sec:1sim} and gives an overview on existing machine learning (ML) approaches to derive digital twins for faster-than-real-time RS predictions in \autoref{sec:1ml}.
Section~\ref{sec:2} presents the employed thermo-mechanical simulation model, and \autoref{sec:3} introduces the methodological approach to surrogate modeling with NODEs and autoencoders. The results are discussed in \autoref{sec:4results} before a summary, and prospective future steps are given in \autoref{sec:summary}.

\subsection{Formation of residual stresses}
\label{sec:1rs}
In general, three different types of RS can be distinguished in DED: In addition to the significant retention of macro stresses after production (type 1), stresses at grain boundaries (type 2) and lattice defects such as voids (type 3) are also considered in the literature \cite{chen_review_2022, bartlett_overview_2019}. 
Several mechanisms can be responsible for the development of type 1 RS, mainly the temperature gradient mechanism and the cooling mechanism \cite{chen_review_2022, bartlett_overview_2019}. In the former, the expansion of the newly forming layer is prevented by the connection with colder, deeper and surrounding layers. This results in compressive stresses in the new layer and tensile stresses in the surrounding area. If the thermal stresses exceed the yield point, this results in plastic deformation. 

In the subsequent cooling mechanism, the effects occur with the opposite sign. 
New layers cannot contract unhindered. Tensile stresses occur here. 
To compensate for this, a tensile stress forms at the component level near the substrate plate, resulting in a tensile-compression-tensile distribution in the direction of build-up \cite{bastola_review_2023, xie_review_2022}. The phase transition in the solid state, for example, observed in the martensite formation of H13 tool steel, is a third mechanism that requires consideration only for certain raw material~\cite{chen_review_2022}. It induces additional strains that can even reverse the sign of the RS distributions.

When developing RS, the individual process and material parameters must also be taken into account~\cite{chen_review_2022}. 
The material parameters have a decisive influence on the development of RS in the context of the temperature gradient mechanism. 
The comparatively low thermal conductivities and high yield and breaking limits make alloys such as Inconel 718 and Ti-6Al-4V particularly susceptible to high RS. \cite{mukherjee_residual_2017} states that the RS can be significantly minimized by reducing the layer thickness. Inconel 718 parts are more susceptible to delamination than Ti-6Al-4V parts as they are exposed to higher RS compared to their yield strength. The general relationships between individual process parameters and the development of RS have so far been partially investigated in the literature \cite{bartlett_overview_2019}. In particular, laser power, deposition rate, vector length and path should be mentioned here. According to a literature review by \cite{chen_review_2022}, there appear to be path patterns that can reduce RS on simple test components~\cite{ren_optimized_2019, robinson_scan_2018}. The island method, the regular repetition of smaller patterns, is one approach to mitigate RS in single-layered experiments~\cite{chen_review_2022, wu_residual_2014}.

\subsection{Simulation of residual stresses}
\label{sec:1sim}
The RS problem motivates the development of various process models -- especially simulations -- and control strategies based on them to compensate for process disturbances and optimize deposition accuracy. 
There are already several approaches for simulating the thermo-mechanical formation of RS. While the focus of the following state of the art is on the DED process, noteworthy approaches from the powder-based DED and powder bed fusion are also listed. The aim is to shed light on the state of the art for the simulation of ES in the interplay between the accuracy of the physical description and the calculation speed.

% commercial packages
Several works employ commercial software packages to investigate AM processes concerning RS.
\cite{ding_thermomechanical_2011} simulate the thermo-mechanical effects of line depositions in Abaqus. 
Four lines of 500 mm each require 75 h calculation time for transient consideration (no specification of CPU resources). Deflection and RS showed good agreement with experimental values. \cite{yang_efficient_2021} manage to reduce the calculation time of the same case to 3:15 h with the help of semi-analytical approaches and mesh coarsening. \cite{sun_residual_2021} carry out similar simulations and their own experiments. Other works implement a thermo-mechanical model in Abaqus \cite{lee_effect_2019,nycz_stress_2021}. 
\cite{nycz_stress_2021} achieved a calculation time of 20--72 h in line printing experiments. 
Especially in the initial phase, a time step of 1 s was necessary to achieve better agreement with neutron diffraction measurements. 
\cite{lee_effect_2019} validated their model using temperature and distortion measurements of the substrate plate. Shorter interlayer cooling phases and bidirectional line application patterns reduced the distortions and RS, also reported in \cite{hu_fea_2023} for powder-based DED simulations of a TB18 titanium alloy. 
\cite{nain_dedthesis_2022} achieved acceleration factors of 5--10 with an acceptable loss of accuracy (10\% error in temperatures) by using a larger equivalent heat source. However, a wall of 50 layers still requires about 2.5--7.5~h plus 5--10~h of simulation time for the thermal and mechanical model (Intel Xeon W-2275, 14 cores). \cite{li_predictive_2021} achieved a speed-up of 19--48 h for a DED simulation in ANSYS Mechanical by simultaneously calculating two layers.
 \cite{biegler_distortions_2018, biegler_FEA_2018} compare deformations for components with their thermo-mechanical model in simucat.welding 7.1 using in-situ 3D digital image correlation (DIC). The calculation time for 30 layers was between 8.8--148.2~h on 16 CPU cores, depending on the mesh resolution. 
 \cite{ren_optimized_2019} investigated the effects on RS of two different zig-zag patterns in a single-layer thermo-mechanical model in MATLAB and COMSOL with stepwise element activation. The calculation took 105.5 h on an Intel Xeon E5-1630 v3 (4 cores at 3.7 GHz), while the temperature had a maximum relative error of 10\%. Lower RS and deflections occurred for application lines along the shorter side of the printed parallelepiped. The modified inherent strain method (ISM) by \cite{liang_inherent_2018} is very popular in the simulation community for efficient RS calculation. In addition to the thermal strains of the original method, the cumulative contribution of the developing elastic strain due to the shrinkage of the upper layers during the cooling process was also taken into account in the calculation of the inherent strains. Lately, different scanning patterns could also be considered in the homogenization for the ISM \cite{liang_scanning_2021}.
 
% acc with own codes
Some works implemented their own code to simulate the DED process.
\cite{huang_stress_2018} achieved a calculation time of 1.8 h for depositing four layers of a line structure using a GPU code developed in-house.
\cite{ou_control_2020} developed a thermo-fluid-mechanical detail-model of the application in their own FORTRAN code to accurately predict the application of a line in shape and temperature (10 min calculation time on Intel i7 2.8 GHz). 
Subsequently, process parameters were optimized using the simulation in an offline approach \cite{ou_control_2020}. 
\cite{ali_residual_2022, ali_residual_2023} developed a thermo-elastoplastic model for arc welding and distinguished solid and molten regions using the phase field method. 

% deposition
There are multiple ways to model the material deposition.
One approach in simulation models is to manipulate the material data, for example, as shown by \cite{peyre_numerical_2008} in the DED process. 
For this purpose, the geometry is completely meshed in advance, and three-dimensional heave-side functions are used to assign the smallest possible values for thermal conductivity and thermal capacity to unprinted areas such that these areas do not significantly influence the result (also termed \emph{quiet} or \emph{penalization element method}). Nowadays, this approach is often replaced by the \emph{inactive element method}, for example in \cite{nain_dedthesis_2022}, or hybrid approaches, e.g. \cite{denlinger_thermomechanical_2017}. 
\cite{pirch_process_2017, pirch_residual_2018} simulated the material application of a layer with a proprietary thermo-mechanical model with mesh movement. 
Similarly, modern Arbitrary Lagrangian Eulerian (ALE) mesh morphing approaches are used in \cite{wirth_predictive_2018} to model the material application. ALE methods are mainly found in detailed models of the melt pool to provide insights into single deposition paths, for example, into the expected widths and heights, and are rather unsuitable for simulations at the component level. For example, \cite{hafiychuk_FEA_2024} implemented an ALE mesh morphing approach to realize the application of a few lines.

%semi analytical
Some research aims to accelerate the predition time with analytical or semi-analytical approaches. 
\cite{huang_rapid_2019} found analytical correlations to predict temperatures during applying individual lines in the powder-based DED process in real time. 
However, the effects of convection and radiation were not taken into account. 
\cite{weisz_phase_2020} developed a semi-analytical solution model of the energy conservation equation in Scilab, which neglects heat flows tangentially to the heat intake direction. Thus, it can predict the temperature distribution in a layer-by-layer fashion (2.6\% error, 20 min calculation time for 100 layers on 7 cores at 2.7 GHz each). 
The prediction of component warpage was later validated using DIC methods \cite{weisz_residual_2022}. 
The elastoplastic material behavior of 316L was modeled using the von Mises flow rule and kinematic hardening. 
\cite{khan_rapid_2022} developed an analytical formula for predicting stress curves in simple geometries. 

% gap: speed for online prediction 
Based on our research in the literature reviews \cite{chen_review_2022, bastola_review_2023, poggi_sota_2022, wei_mechanistic_2021} and beyond, simulation models have not yet been used for the online prediction of RS. 
In most cases, these approaches are either too complex, such that the computation times are far from real-time predictions or fast but inaccurate. 
This section demonstrates that without employing surrogate models, the simulation acceleration factors of about $10^4$ compared to real time -- as required for the online correction of RS~\cite{kannapinn_digital_2022, kannapinn_digital_2023, kannapinn_twinlab_2024} -- cannot be achieved with low CPU resources and sufficient accuracy. 

\subsection{ML surrogates and digital twin concepts for RS design}
\label{sec:1ml}
Employing ML surrogates or reduced order models (ROMs) based on highly accurate physical models can remedy this. The literature review \cite{wei_mechanistic_2021} reveals that in the existing AM literature, ROMs are mainly understood as models with reduced simulated physics and lower temporal or spatial resolution. 
There are some purely data-based approaches to train fast surrogate models. According to recent literature reviews by \cite{era_MLDED_2023, wei_mechanistic_2021} on ML in the field of DED processes, ML methods were mainly used for the offline categorization of process anomalies and component defects. For this purpose, quality predictions are derived from sensor data \cite{khanzadeh_porosity_2018, chen_sensor_2023, mi_monitoring_2023}. These static mappings between sensor data and quality parameters are not derived from first principles and can be classified as purely phenomenological.

According to the following literature reviews, only a few works are dedicated to online DED process or quality optimization, although the need is clearly formulated. The following current conclusion by \cite{era_MLDED_2023} on the state of the art says it all: 

\begin{displayquote}Therefore, it is evident from the available literature that the current technology is still not capable of enabling process monitoring and online optimization in DED, making this problem class the most potential research field for the future. Online optimization with preventive maintenance can drastically change material waste, repetitive post-processing, and overall print quality in DED.
\end{displayquote}
According to \cite{bartlett_overview_2019}, there are still no online approaches to measure, predict or control RS:

\begin{displayquote}To better understand residual stress development during additive manufacturing moving forward, we conclude that non-invasive in situ monitoring methods are necessary [...]. While significant research has been conducted on in-process measurements of temperature or irregularities with regards to porosity or microstructure [...], very little has been reported to measure RS in-process […]. The development of effective full-field in-process RS measurements may result in achievable real-time control strategies to either eliminate RS or induce intentional anisotropy and optimized properties for specific applications […] we envision as the ultimate goal of work in this field: residual stress control and optimization, process qualification. An appealing route to achieving this goal is the utilization of machine learning (ML) techniques.
\end{displayquote}

Overall, no significant ROM or ML approaches to date allow RS in DED to be predicted with sufficient accuracy, reliability and real-time capability to control the process online. The following section lists the few noteworthy works in the context of this study: \cite{kozjek_datadriven_2022} uses a random forest regressor to predict the temperature distribution of the next layer based on recorded on-axis pyrometry data. The approach thus belongs to the class of N-step-ahead regressors and cannot be used as an independent simulation model. \cite{dharmadhikari_reinforcement_2023} found optimal parameter ranges for laser power and feed rate using a reinforcement learning algorithm. However, the predictions are based on the simplified Eager--Tsai model for a DED process. \cite{brown_ROM_2023} use the Gaussian process latent space dynamics identification (GPLaSDI) model to predict the temperature distribution of simple, synthetic DED line jobs. %, but do not demonstrate any interaction with a real system. 
\cite{kushwaha_advanced_2024} trained a parametric deep operator network to predict sequentially coupled thermo-mechanical field data from a simplified DED process. Noteworthy, a maximum relative $L^2$ error of 9.27\% was achieved to predict the von Mises equivalent stress. However, the focus was less on the precise modeling of realistic DED processes, as the time dependence of the heat conduction problem was not considered. % -- and also RS control.
%The processing of different geometries shows the hardware requirements for modern digital twin research in the DED field: High performance computers with 64 computing cores for around 5000 ABAQUS simulations and NVIDIA A100 modules with large GPU memory for subsequent machine learning were necessary. 
An approach very similar to \cite{kushwaha_advanced_2024} is also pursued by \cite{karkaria_digital_2024}, who combine long short-term memory (LSTM) surrogates with Bayesian methods to predict the distribution of component temperatures in individual points. However, the prediction of the RS and its control in real time is missing. \cite{ren_thermal_2020} introduced a promising concept for the ML of deposition paths. For this purpose, each layer of the CAD model of the part is divided into unit cells. Arbitrarily shaped unit cells were thermally simulated in advance with different deposition patterns and learned with a combined ansatz of recurrent neural networks (RNN) und deep neural networks (DNN). Some 100 cases could be pre-simulated in 50 days using two high-performance clusters. \cite{vohra_surrogate_2020} derived a model reduction from thermo-mechanical simulations of an electron beam process using principal components analysis (PCA), subsequently allowing the prediction of the component's stress peaks. 

In summary, the literature's relevant, time-dependent ML models of the DED process are mainly limited to RNNs and LSTMs. These models require a lot of training data from computationally expensive simulations and have only limited predictive capabilities. Thus, there has been no research on data-driven, physics-based ML methods for accurate, reliable and data-efficient surrogate modeling and RS prediction in the DED process.

\section{Thermo-mechanical simulation of residual stress formation}
\label{sec:2}

\begin{figure}[b!]
    \centering
        \includegraphics[width=0.7\textwidth,trim=0 0 10 0, clip]{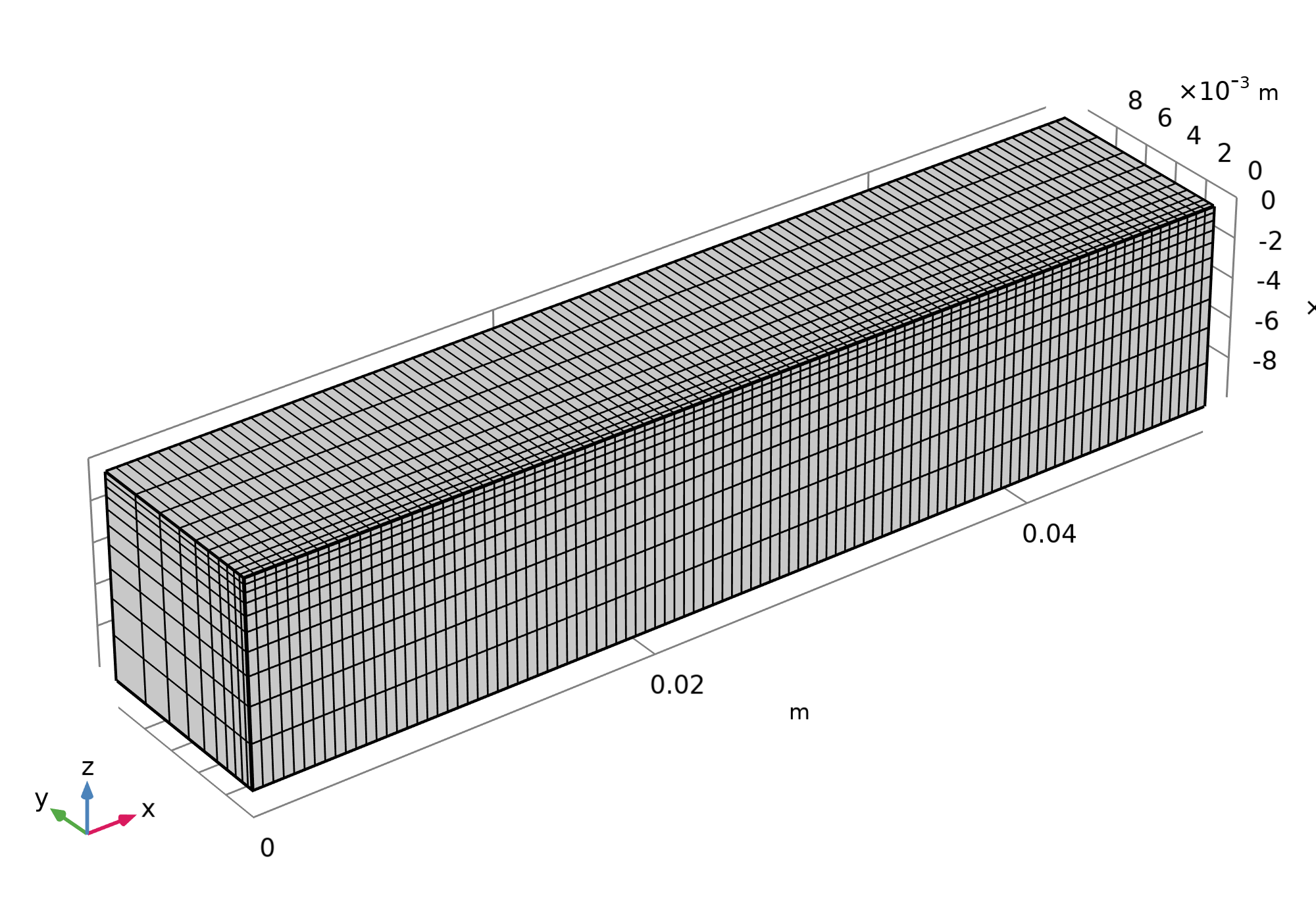}
        \caption{Mesh and dimensions of the simulation domain.}
    \label{fig:MeshDomain}
\end{figure}

A simulation model of the DED process was set up in COMSOL Multiphysics 6.2 employing a solid-state thermo-elasto-plasticity model, similar to \cite{mukherjee_residual_2017, nain_dedthesis_2022}. 
One laser line scan of the DED process over a block of Inconel 718 with dimensions $5 \times 1 \times 1~\unit{cm}$ is simulated, see \autoref{fig:MeshDomain} for a representation.  
For computational efficiency, the plane $y=0$ is considered a symmetry plane. 
Temperatures are calculated with the solid-state heat conduction equation
\begin{align}
& \rho(T) c_p(T) \frac{\partial T}{\partial t}=\frac{\partial}{\partial x_i}\left(k(T) \frac{\partial T}{\partial x_i}\right),
\end{align}
where $T$ is temperature, $t$ is time, $\rho$ is density, and $c_p$ is heat capacity at constant pressure. The temperature-dependent material properties can be found in \autoref{tab:mat2}, whereas the temperature-dependent 
thermal conductivity
\begin{equation}
 k(T) = 0.56+ \num{2.9E-2} \, T - \num{7E-6} \, T^2
\end{equation}
is taken from \cite{mukherjee_residual_2017}. 
\begin{table}[b!]
\addtolength{\tabcolsep}{+2.1pt}
\centering
\caption{Temperature-dependent material properties of Inconel 718 \cite{romano_laser_2016, kamara_numerical_2011}.}
\begin{tabular}{rllllll}
\toprule
$T$/K & $c_p$/\unit{J.kg^{-1}.K^{-1}} & $\rho$/\unit{kg.m^{-3}} & $\epsilon$/- & $E$/GPa & $\sigma_\text{y}$/MPa & $\alpha \times 10^{-6}$/\unit{K^{-1}} \\ \midrule
298             & 435                   & 8190                                 & 0.539      & 200   & 1125  & 12.20 \\
373             & 455                   & 8160                                 & 0.533      &       &       &       \\
473             & 479                   & 8118                                 & 0.533      &       &       & 14.36 \\
573             & 497                   & 8079                                 & 0.534      &       &       & 14.90 \\
673             & 515                   & 8040                                 & 0.534      & 178   &       & 15.43 \\
773             & 427                   & 8001                                 & 0.535      &       & 1020  &       \\
873             & 558                   & 7962                                 & 0.535      & 163   & 965   & 17.45 \\
973             & 568                   & 7925                                 & 0.536      &       &       &       \\
1073            & 680                   & 7884                                 & 0.536      &       & 800   & 18.34 \\
1173            & 640                   & 7845                                 & 0.537      & 139   &       &       \\
1273            & 620                   & 7806                                 & 0.537      &       &       &       \\
1373            & 640                   & 7767                                 & 0.538      & 99    &       &       \\
1443            & 650                   & 7727                                 & 0.538      &       &       &       \\
1533            &                       &                                      &            &       & 25    &       \\
1609            & 720                   & 7400                                 & 0.329      & 1     & 1     &       \\
1673            &                   & 7340                                 & 0.332      &       &       &       \\
1773            &                    & 7250                                 & 0.337      &       &       &       \\
1873            &                  & 7160                                 & 0.341      &       &       &      \\
\bottomrule 
\end{tabular}
\label{tab:mat2}
\end{table}
To account for the latent heat $L_\mathrm{m} = \SI{230E3}{J/kg}$ required for the transition from solid to liquid, the heat capacity is modified in the temperature range between solidus $T_s = \SI{1533}{K}$ and liquidus temperature $T_l = \SI{1609}{K}$ following~\cite{hafiychuk_FEA_2024} as
\begin{equation}
    c_p^*(T)=c_p(T)+L_\text{m} \frac{1}{\sqrt{\pi \Delta T^2}} \exp \left(-\frac{\left(T-T_{\mathrm{m}}\right)^2}{\Delta T^2}\right)\,,
\end{equation}
where $\Delta T = T_\mathrm{l} - T_\mathrm{s}$ and $T_{\mathrm{m}} = \left( T_\mathrm{l} - T_\mathrm{s} \right)/2$.
Increased mixing within the melted zone due to the Marangoni effect is accounted for with a modified thermal conductivity~\cite{ren_optimized_2019}
\begin{align}
    k^*(T) = 2.5 k(T) \quad \forall T>T_{\text{m}} \,.
\end{align}

The convective and radiative thermal boundary conditions read for all boundaries except for the symmetry plane
\begin{align}
& Q_{\mathrm{conv}}+Q_{\mathrm{rad}}=-h_{\mathrm{conv}}\left(T_\text{surf}-T_{\mathrm{amb}}\right)-\epsilon \sigma\left(T_\text{surf}^4-T_{\mathrm{amb}}^4\right),
\end{align}
where the convective heat transfer coefficient is $h_\text{conv} = \SI{10}{W.m^{-2}.K^{-1}}$, the temperature-dependent emissivity $\epsilon$ can be found in \autoref{tab:mat2}, $\sigma$ is the Stefan--Boltzmann constant and the ambient temperature is $T_\text{amb} = \SI{293.15}{K}$. The surface temperature $T_\text{surf}$ originates from the current solution of $T$ at the boundary.

The heat source of the laser was modeled with a Goldak double-ellipsoid model \cite{golak_fea_1984} 
\begin{align}
Q(x_i, t)=\frac{6 \sqrt{3} \eta P}{a_{\mathrm{f/r}} b c \pi \sqrt{\pi}} \exp \left(-\left(\frac{3(x_1+V t)^2}{a_{\mathrm{fr}}{ }^2}+\frac{3 x_2^2}{b^2}+\frac{3 x_3^2}{c^2}\right)\right) F_{\mathrm{f/r}},
\end{align}
which can be considered a standard approach in modeling laser-based additive manufacturing. The corresponding model constants are in \autoref{tab:goldak}.
\begin{table}[t!]
\caption{Goldak double-ellipsoid model constants.}
\centering
\begin{tabular}{l c l}
\toprule
Goldak parameter                      & ~~~~Symbol~~~~                    & Value  \\ \midrule
Absorption efficiency                     & $\eta$                         & 0.4 \cite{prom_comprehensive_2017}   \\ 
Laser power                             & $P$                          & \SI{1000}{W}  \\ 
Laser velocity                            & $V$                          & \SI{15}{mm.s^{-1}}  \\ 
Front ellipsoid length                & $a_f$                     & 3 mm \\ 
Rear ellipsoid length                 & $a_r$                     & 8 mm \\ 
Ellipsoid width                       & $b$                         & 3 mm \\ 
Ellipsoid depth                       & $c$                         & 3 mm \\ 
Weighing fraction for front ellipsoid & $\mathrm{F}_{\mathrm{f}}$ & 0.67    \\ 
Weighing fraction for rear ellipsoid  & $\mathrm{F}_{\mathrm{r}}$ & 1.33   \\\bottomrule
\end{tabular}
\label{tab:goldak}
\end{table}

Assuming small deformations and strains, the quasi-static thermo-elasto-plasticity model reads
\begin{align}
\vect{\nabla}_x \cdot \vect{\sigma}&=\vect{0} \,,\\
\vect{\varepsilon}&=\vect{\varepsilon}_{\mathrm{el}}+\vect{\varepsilon}_{\mathrm{pl}}+\vect{\varepsilon}_{\mathrm{th}} 
= \tfrac{1}{2}\big(\vect{\nabla}_x \vect{u}+ \vect{\nabla}_x \vect{u}^T\big)\,,\\
\vect{\varepsilon}_{\mathrm{el}}&=\frac{(1+\nu)}{E} \vect{\sigma}-\frac{\nu}{E} \operatorname{tr}(\vect{\sigma})\,, \\
\vect{\varepsilon}_{\mathrm{th}}&=\alpha\left(T-T_{\mathrm{amb}}\right) \vect{I}\,,
\end{align}
where $\vect{\sigma}$ is the Cauchy stress tensor, $\vect{\varepsilon}_{\mathrm{el}}$ are the elastic, $\vect{\varepsilon}_{\mathrm{pl}}$ are the plastic, and $\vect{\varepsilon}_{\mathrm{th}}$ are the thermal strains, and $\vect{u}$ is the displacement field. $E$ and $\nu=0.28$ denote Young's modulus and Poisson's ratio, respectively. The volumetric expansion coefficient $\alpha$ and the identity matrix $\vect{I}$ account for isotropic thermal expansion. 
The relevant material properties are taken from \cite{romano_laser_2016,kamara_numerical_2011} and can be found in \autoref{tab:mat2}.

The von Mises yield criterion is applied
$$
F_{\mathrm{y}}=\sigma_{\mathrm{m}}-\sigma_\text{y}\!\left(\varepsilon_{\mathrm{pe}}, T\right)=0\,,
$$
where $\sigma_\text{m}$ and $\varepsilon_\text{pe}$ are the von Mises equivalent stress and plastic strain and $\sigma_\text{y}$ is the temperature-dependent yield stress, as given in \autoref{tab:mat2}. 
All temperature-dependent parameters are interpolated with piece-wise cubic splines and extrapolated constantly.
Bilinear, isotropic hardening with a tangent modulus of \SI{2}{GPa} is estimated based on \cite{soffel_interface_2020}.
Moreover, the modeling of a phase transition in the solid state is omitted, as, according to \cite{denlinger_stress_2016, bartlett_overview_2019}, this effect is not relevant for the correct modeling of the RS in Inconel materials in comparison to Ti-6Al-4V for example. 
At $z=0$, a roller boundary condition is employed, while point $[0,0,0]$ is fixed in all directions for a well-defined statical problem.

\section{Compressed latent space dynamics surrogate with neural ODEs and autoencoders}
\label{sec:3}

The foundation of successful digital twin derivation lies in the physics-based modeling of the additive manufacturing process, as outlined above. The RS in the manufactured component are to be predicted during the ongoing DED process based on the available current process parameters. 
However, the need to provide faster-than-real-time replications of these mappings through simulations presents a challenge, particularly as the complexity and computational cost of multi-physical simulation models increase. 
In the context of model predictive control, simulation models are often called \numrange{10}{100} times by gradient-based optimizers to determine the optimum further course of the process \cite{kannapinn_digital_2023}.
It is, therefore, necessary to accelerate the dynamic prediction model compared to real time \cite{kannapinn_digital_2022, kannapinn_twinlab_2024}, without significant loss of accuracy. 
To achieve this, approaches that combine the flexibility and prediction capabilities of machine learning with the rigour and generalizability of physics-based modeling concepts must be developed \cite{watson2024}.
This study addresses the challenge above by developing a surrogate based on an efficient ML reduced-order modeling methodology, utilizing autoencoders for spatial compression and NODEs for latent space dynamics identification (LaSDI).

\subsection{Neural ordinary differential equations} 
\label{sec:nodes}

Recently, hybrid approaches have emerged that combine identifying system dynamics with machine-learning techniques~\cite{rom_bennerMOR1}. 
Here, NODEs \cite{chen_node_2018} are employed to reproduce the transient output of temperatures and stresses by solving an ODE over multiple time steps:
\begin{align}\label{eq:DynROMODE}
%\frac{\partial}{\partial t} \begin{bmatrix} \vect{X} \\ \vect{I} \end{bmatrix} &= \vect{f} \left( \begin{bmatrix} \vect{X}\\ \vect{I}\end{bmatrix}, G \right),\\
    \frac{\partial}{\partial t} \vect{z} &= \vect{f}(\vect{z},\vect{v}) \,,\\
    \vect{z}(t=0) &= \vect{z}_0  \,, %= AE(x_0)
\end{align}
where $\vect{z} \in  \mathbb{R}^{n_l}$ is a vector containing field variables, $\vect{z}_0 $ contains their respective initial condition and $\vect{v} \in  \mathbb{R}^{m}  $  is a vector of external inputs to the ODE, e.g., temperatures for the NODE predicting stresses.
As neural networks are known to be universal nonlinear function approximators~\cite{rom_Hornik1991MLP,rom_goodfellow16DL}, a feed-forward neural network represents the right-hand side operator $\vect{f}$ of the ODE, which inspired the name of the approach. 
Recent publications indicate that NODEs outperform RNNs or tree-based algorithms~\cite{rom_Lu21node-vs-lstm-pharma,rom_pepe22node-vs-lstm-battery}. 
The success of NODEs is conjectured to be the learning of underlying dynamics instead of just input-to-output relations~\cite{rom_neuralODEblog,rom_pepe22node-vs-lstm-battery}.
NODEs seem promising for predicting the system dynamics of RS. To the best of the authors' knowledge, they have not yet been investigated in the context of AM.

\subsection{Autoencoders}
\label{sec:aes}

Replicating all $n_s$ nodal temperatures or stresses over all $n_t$ time steps of the full-order model would render a NODE considerably high-dimensional and inefficient to train~\cite{kannapinn_digital_2023}. 
However, many complex systems can be described by dominant low-dimensional patterns without considerable loss of accuracy. 
The singular value decomposition (SVD) would be an obvious choice to find a low-rank approximation of the dynamic problem, as it is executed in the proper orthogonal decomposition method~\cite{rom_brunton22databook}.
However, the SVD provides only a linear decomposition of the problem and is reported to fail providing a low-rank approximation especially for convective problems. In contrast, autoencoders show for a good nonlinear mapping to a latent space~\cite{brown_ROM_2023}.

\begin{figure}[tbp]
\centering
\begin{tikzpicture}[x=1cm,y=1cm, scale=0.7]

    \normalsize
    % \large
    
    \def\Ysep{2.5}

    % Input layer
    \node[node inout large] (PK11) at (0,  0.5*\Ysep) {$y(t_0)$};
    \node[below] at (PK11.south) {\small Initial conditon};
    \node[node inout large] (T) at (0, -0.5*\Ysep) {$u(t)$};
    
    \node (FOM) at (0,  1*\Ysep) {Field input};

    % Encoding layer
    \def\Xoff{2.5}
    \node[node encoder] (AE_enocde_PK11) at ([xshift=\Xoff cm] PK11) {};
    \node at ([xshift=\Xoff cm] PK11) {$\phi_y$};
    \node[node encoder] (AE_enocde_T) at ([xshift=\Xoff cm] T) {};
    \node at ([xshift=\Xoff cm] T) {$\psi_u$};

    \draw[connect arrow] (PK11) -- (AE_enocde_PK11);
    \draw[connect arrow] (T) -- (AE_enocde_T);

    % ODE Layer
    \node[node layer] (ODE) at ([xshift=1.4*\Xoff cm] AE_enocde_PK11) {$\,\dot z = f(z, v)\,$};

    \draw[connect arrow] (AE_enocde_PK11) -- (ODE) node[pos=0.5,above] {$z_{0}$};
    \draw[connect arrow] (AE_enocde_T) -| (ODE) node[pos=0.8, right] {$v(t)$};
    
    % Integration layer
    \node[node layer] (Integrator) at ([xshift=1.1*\Xoff cm] ODE) {$\int$};

    \draw[connect arrow] (ODE) -- (Integrator);

    % Decoder layer
    \node[node decoder] (AE_decode_PK11) at ([xshift=\Xoff cm] Integrator) {};
    \node at ([xshift=\Xoff cm] Integrator) {$\phi_y^{-1}$};
    \coordinate (h0) at ($(AE_enocde_PK11)!0.5!(AE_decode_PK11)$);
    \node at (h0 |- FOM) {Latent space};
    
    \draw[connect arrow] (Integrator) -- (AE_decode_PK11) node[midway,above] {$z(t)$};

    % Output layer
    \node[node inout large] (PK11_out) at ([xshift=\Xoff cm] AE_decode_PK11) {$y(t)$};
    \draw[connect arrow] (AE_decode_PK11) -- (PK11_out);
    
    \node at (PK11_out |- FOM) {Field output};
     
  \end{tikzpicture}
\caption{NODE LaSDI machine learning architecture. For predictions of RS, a zero distribution is set as initial condition $y(t_0)$, and temperatures are the external input $u(t)$ to the surrogate. } % \footnotemark \protect \textsuperscript{\thefootnote} 
 \label{fig:nodeLaSDI}
\end{figure}
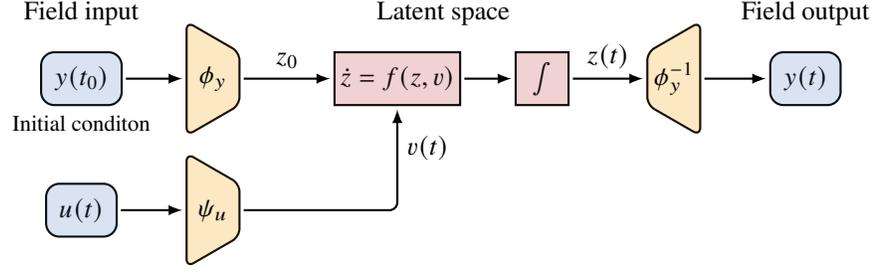

Two autoencoder-based NODE models are evaluated sequentially to predict the stresses over time. The first, denoted $\modela$, maps the heat source field and the initial temperature distribution to the transient temperature field. In turn, this transient temperature field is used with the initial stress distribution as an input to the second model $\modelb$, which finally yields the transient stress field. Since both models are highly similar, we describe a general version $\generalmodel$ predicting generic field data $y$ from an excitation $u$ (i.e., $y=T,~u=Q$ for predicting temperatures from heat sources and $y=\sigma,~u=T$ for predicting RS from temperatures). This general model is illustrated in \autoref{fig:nodeLaSDI}. 
The encoder stage $\phi_y$ maps the initial field data $y(t_0)$ to its latent representation $z_0$. 
Similarly, the encoder stage $\psi_u$ maps the full spatiotemporal input field data $u(t)$ to its latent representation $v(t)$. In the latent space, the NODE then integrates the initial condition $z_0$ over time using $v(t)$ as external input.
Finally, the integrated latent trajectory is mapped back to the full-order output via the decoder stage $\phi_y^{-1}$.

Each model thus consists of two encoders, one decoder and two NODEs. All encoder stages consist of feed-forward neural networks $\phi_y,\psi_u$ with $[n_s, 1024, 128, n_l]$ neurons per layer, including the input layer. Here, $n_s$ denotes the number of spatial points at which the full-order field is captured, and $n_l$ is the dimensionality of the latent space. In the results presented below, the data consisted of $n_s=1836$ spatial points, and $n_l$ is varied from \numrange{1}{10}. The decoder stages $\phi_y^{-1}$ use the same number of neurons per layer as their encoder counterparts but in reversed order. The feed-forward neural networks $f$ of the NODEs have $[2\cdot n_l, 16, 16, n_l]$ neurons per layer, again including the input layer. In all neural networks, the softplus activation function is used throughout the hidden layers. An adaptive step size Runge--Kutta integrator of order five is used for NODE integration. The model parameters are optimized utilizing the adabelief algorithm \cite{zhuang2020}. In addition, adaptive gradient clipping is employed to avoid potential issues with exploding gradients.

\section{Results and discussion}
\label{sec:4results}
This proof-of-concept study investigates whether combining autoencoders and NODEs can predict transient temperature and stress fields of DED additive manufacturing. 
Note that a material deposition is not considered but could be added to the thermo-mechanical simulation model as a next step towards a realistic process simulation.

\subsection{Thermo-mechanical simulation}

The final computational domain of the thermo-mechanical simulation model (compare \autoref{sec:2}) was discretized with hexahedral elements having a height of \SIrange{0.25}{2}{mm} with a refinement towards the heat-affected zone. 
Figure~\ref{fig:MeshDomain} illustrates the computational mesh. 
Generalized Richardson extrapolation~\cite{grid_roache} was employed to estimate the grid-independent solution of temperatures and stresses with two coarser meshes of factor two, respectively. 
The relative discretization errors of \SI{0.0016}{\%} and \SI{0.1392}{\%} compared to the grid-independent solution for temperature and 11-stress component $\sigma$ were calculated.
The system of equations was solved in a segregated manner for energy and momentum conservation. 
Employing quadratic Lagrange elements, the resulting system had \num{2262332} degrees of freedom. 
A deposition of \SI{20}{s} real time was solved with a read-out time step of \SI{0.1}{s}, while the backward differentiation formula (BDF) time stepper of order one employed also smaller and larger time steps, depending on the internal error estimator operating with a default tolerance of \SI{0.1}{\%} relative error. 
One simulation took approximately \SI{10}{min} on 16 cores of a cluster PC with two {Intel Xeon E5-2687W v4 (3.2 GHz) CPUs.}
\begin{figure}[tb]
    \centering
     \begin{subfigure}[]{0.49\columnwidth}
        \centering
        \includegraphics[width=0.99\textwidth]{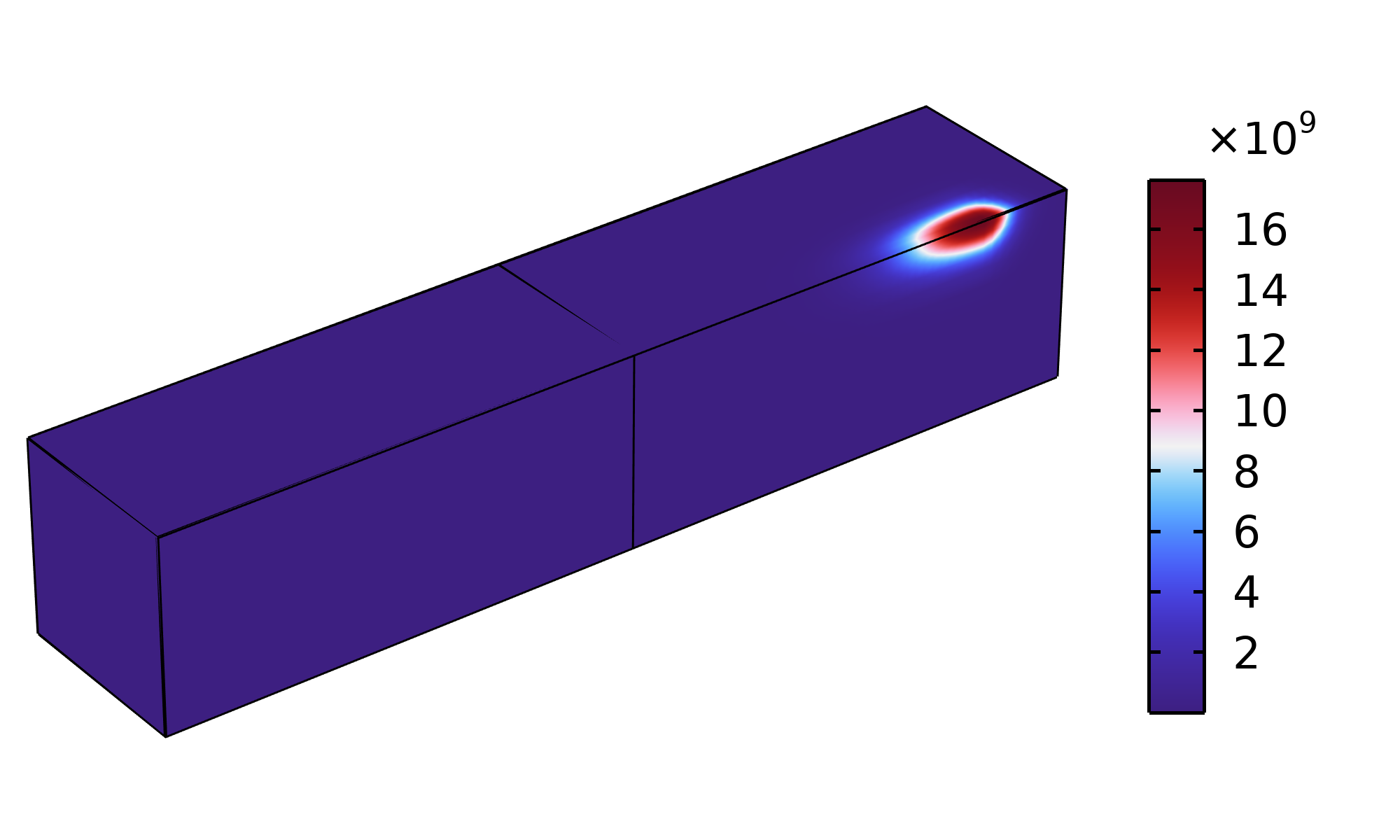}
        \caption{Volumetric heat source Q (\unit{W.m^{-3}}).}\label{fig:after3sQ}
    \end{subfigure}
 \begin{subfigure}[]{0.49\columnwidth}
        \centering
        \includegraphics[width=0.99\textwidth]{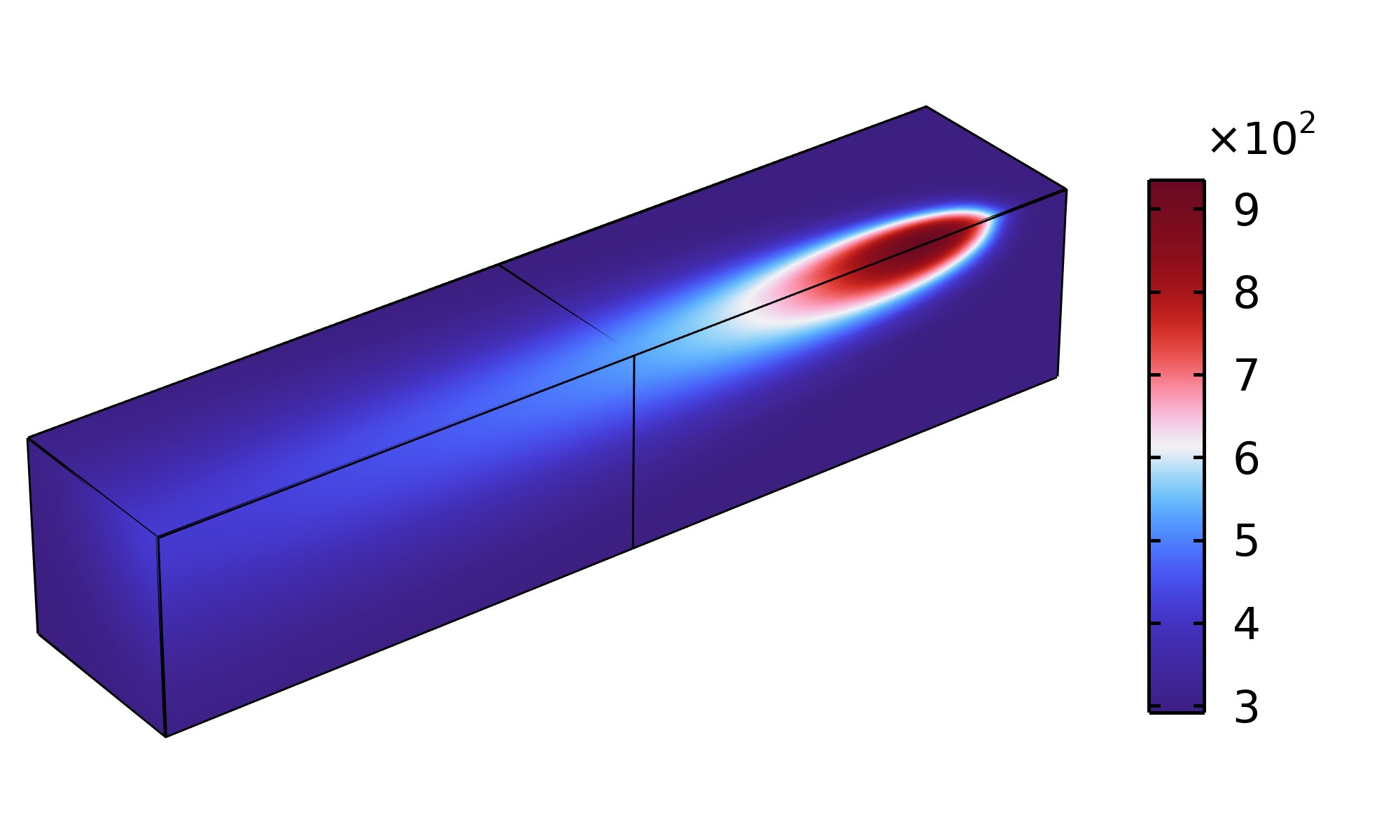}
        \caption{Temperature (K).}\label{fig:after3sT}
    \end{subfigure} 
    \begin{subfigure}[]{0.49\columnwidth}
        \centering
        \includegraphics[width=0.99\textwidth]{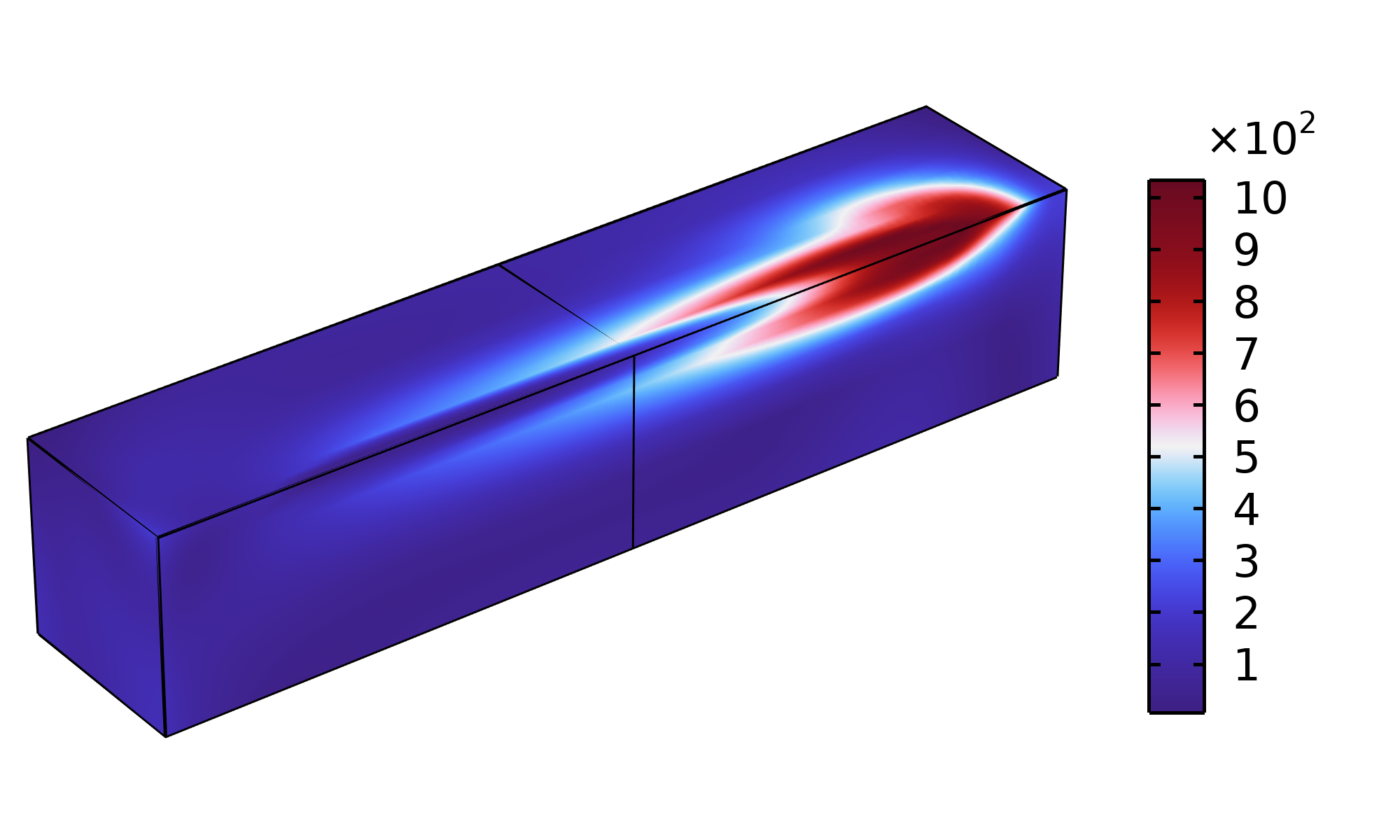}
        \caption{Von Mises stress (MPa).}\label{fig:after3smises}
    \end{subfigure}
    \begin{subfigure}[]{0.49\columnwidth}
        \centering
        \includegraphics[width=0.99\textwidth]{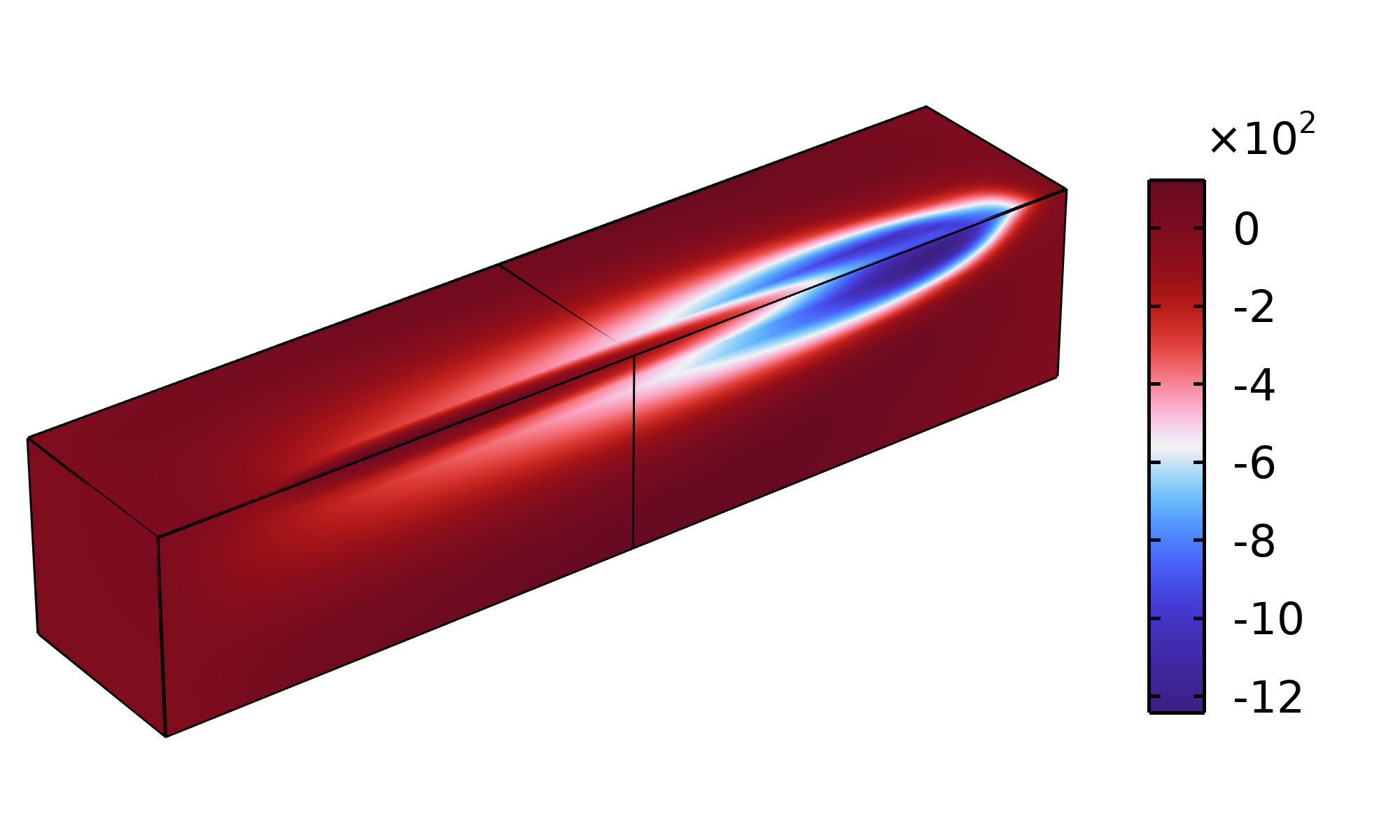}
        \caption{11-stress component of $\vect{\sigma}$ (MPa).}\label{fig:after3ssig11}
    \end{subfigure}
    \caption{Selected simulation results at $t=\SI{3}{s}$.}
    \label{fig:after3sSim}
\end{figure}

Figure~\ref{fig:after3sSim} depicts some illustrative simulation results after 3~s of real time. The heat source $Q$, moving from left to right, induces a longitudinal thermal profile within the slab. The resulting RS form as a thin line (dark blue in \autoref{fig:after3smises} and dark red in \autoref{fig:after3ssig11} at $y=z=0$) right behind the temporary stress field in the direct surrounding of the moving heat source.

\begin{figure}[tb]
    \centering
    \begin{subfigure}[]{0.49\columnwidth}
    \centering
        \includegraphics[width=0.8\textwidth,trim=330 0 370 0, clip]{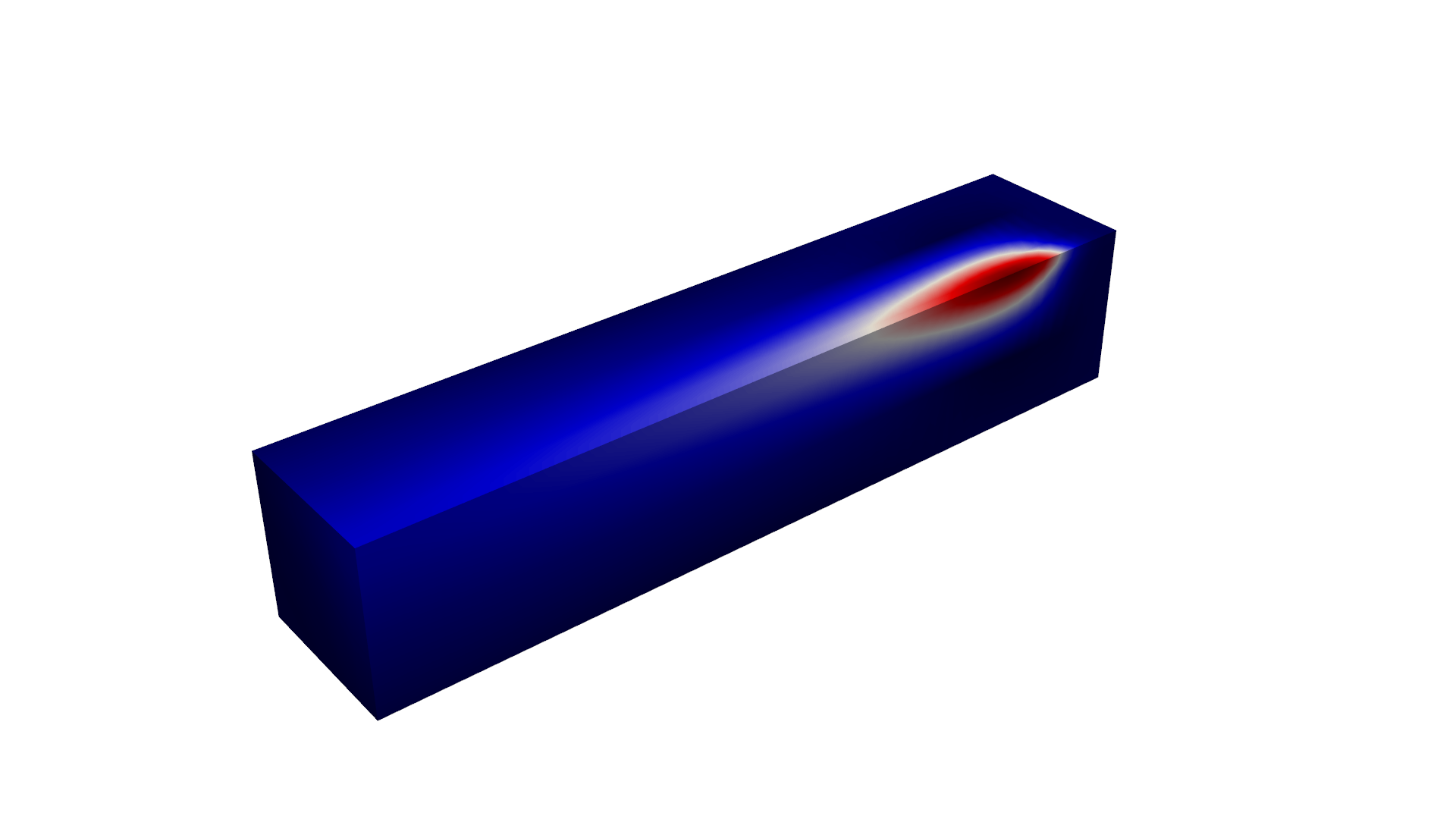}
        \includegraphics[width=0.15\textwidth,trim=0 0 0 0, clip]{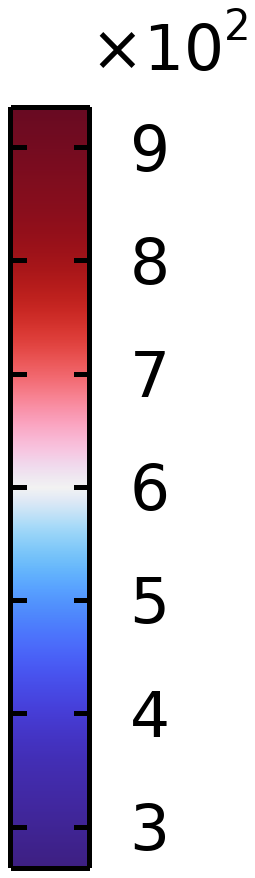}
        \caption{Temperature (K) target $t=\SI{3}{s}$.}\label{fig:after3sa}
    \end{subfigure}
    \begin{subfigure}[]{0.49\columnwidth}
        \centering
        \includegraphics[width=0.8\textwidth,trim=330 0 370 0, clip]{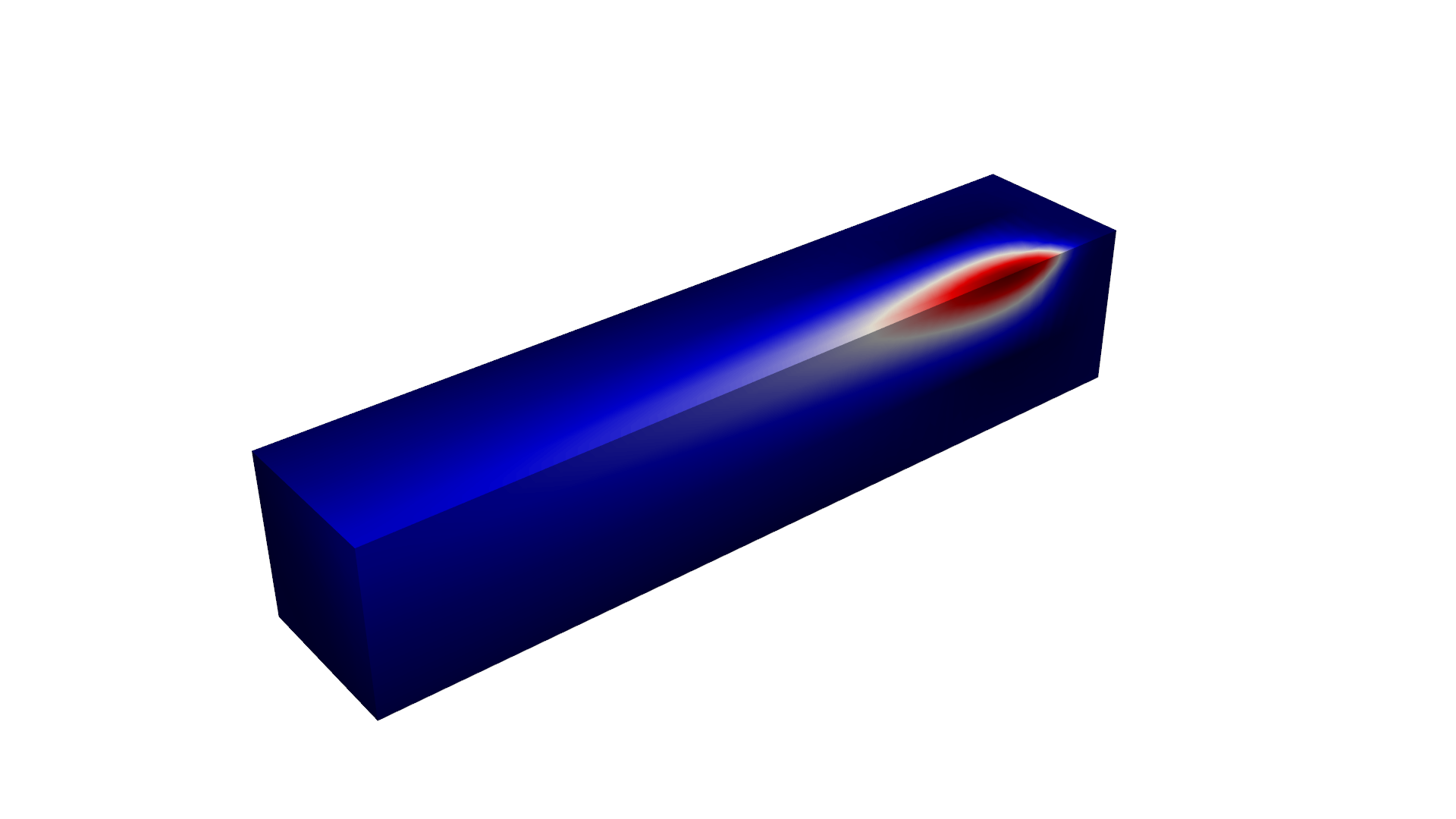}
        \includegraphics[width=0.15\textwidth,trim=0 0 0 0, clip]{graphics/leg_fabian_t3_wave_T}
        \caption{Temperature (K) prediction $t=\SI{3}{s}$.}\label{fig:after3sb}
    \end{subfigure}
    \begin{subfigure}[]{0.49\columnwidth}
        \centering
        \includegraphics[width=0.8\textwidth,trim=330 0 370 0, clip]{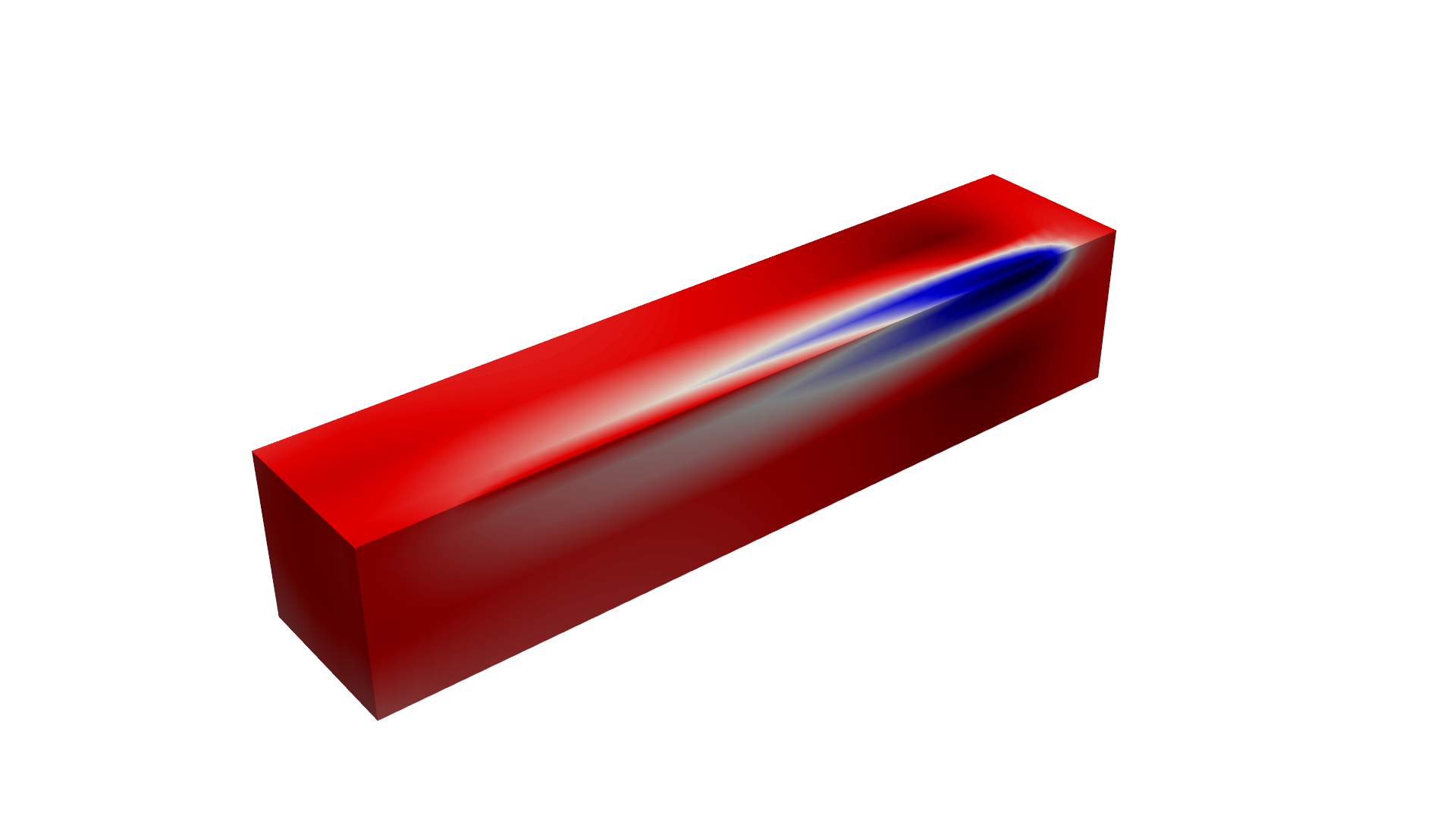}
        \includegraphics[width=0.15\textwidth,trim=0 0 0 0, clip]{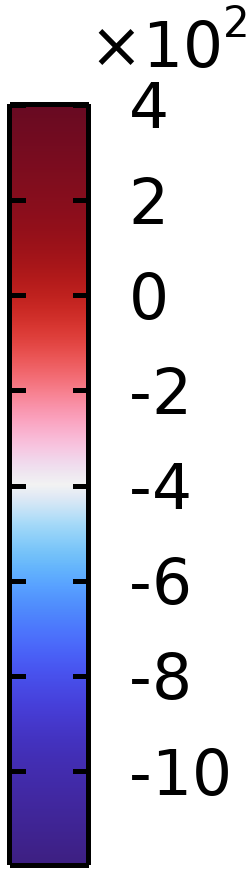}
        \caption{Stress (MPa) target $t=\SI{3}{s}$.}\label{fig:after3sc}
    \end{subfigure}
    \begin{subfigure}[]{0.49\columnwidth}
        \centering
        \includegraphics[width=0.8\textwidth,trim=330 0 370 0, clip]{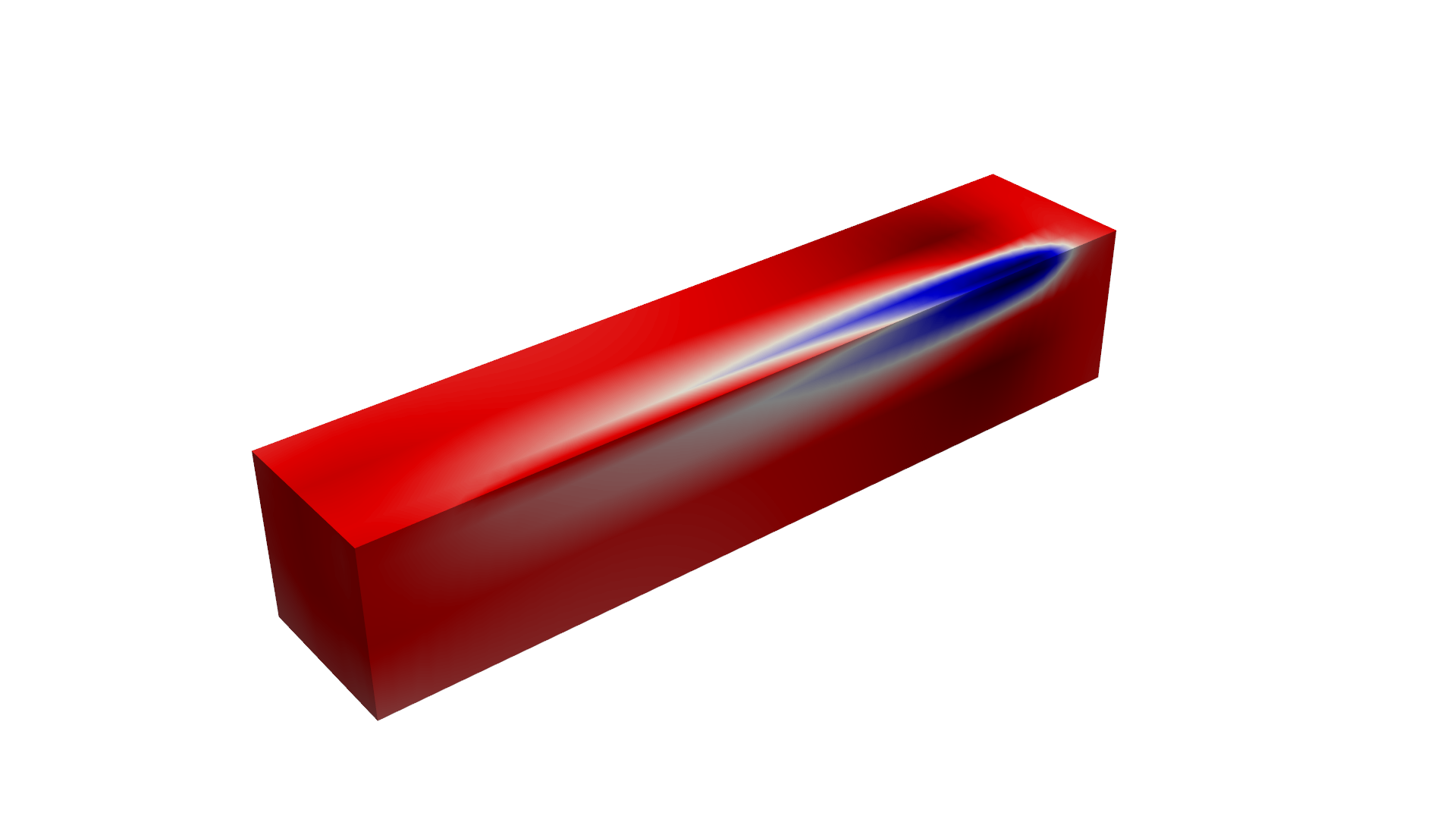}
        \includegraphics[width=0.15\textwidth,trim=0 0 0 0, clip]{graphics/leg_fabian_t3_wave_MPa_pk11}
        \caption{Stress (MPa) prediction $t=\SI{3}{s}$.}\label{fig:after3sd}
    \end{subfigure}
    \begin{subfigure}[]{0.49\columnwidth}
        \centering
        \includegraphics[width=0.8\textwidth,trim=330 0 370 0, clip]{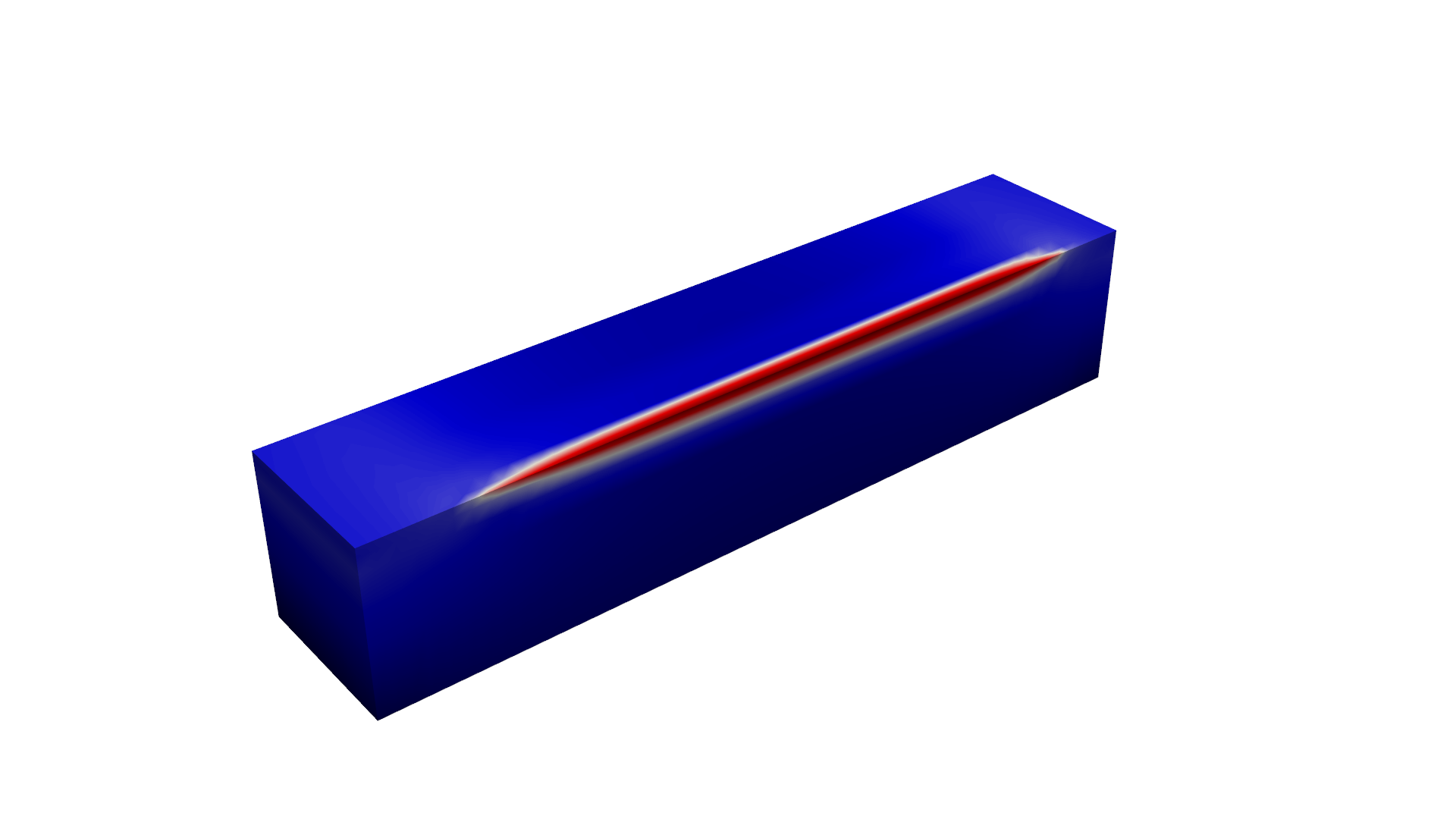}
        \includegraphics[width=0.17\textwidth,trim=0 0 0 0, clip]{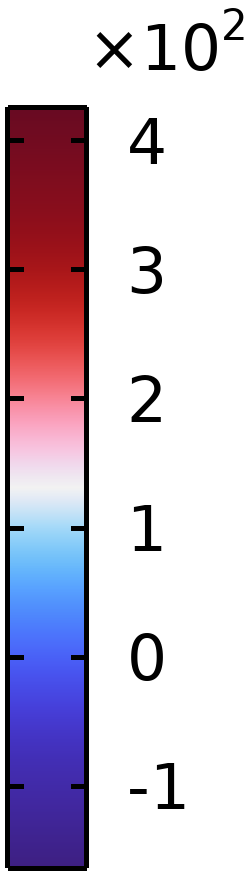}
        \caption{Stress (MPa) target $t=\SI{20}{s}$.}\label{fig:after3se}
    \end{subfigure}
    \begin{subfigure}[]{0.49\columnwidth}
        \centering
        \includegraphics[width=0.8\textwidth,trim=330 0 370 0, clip]{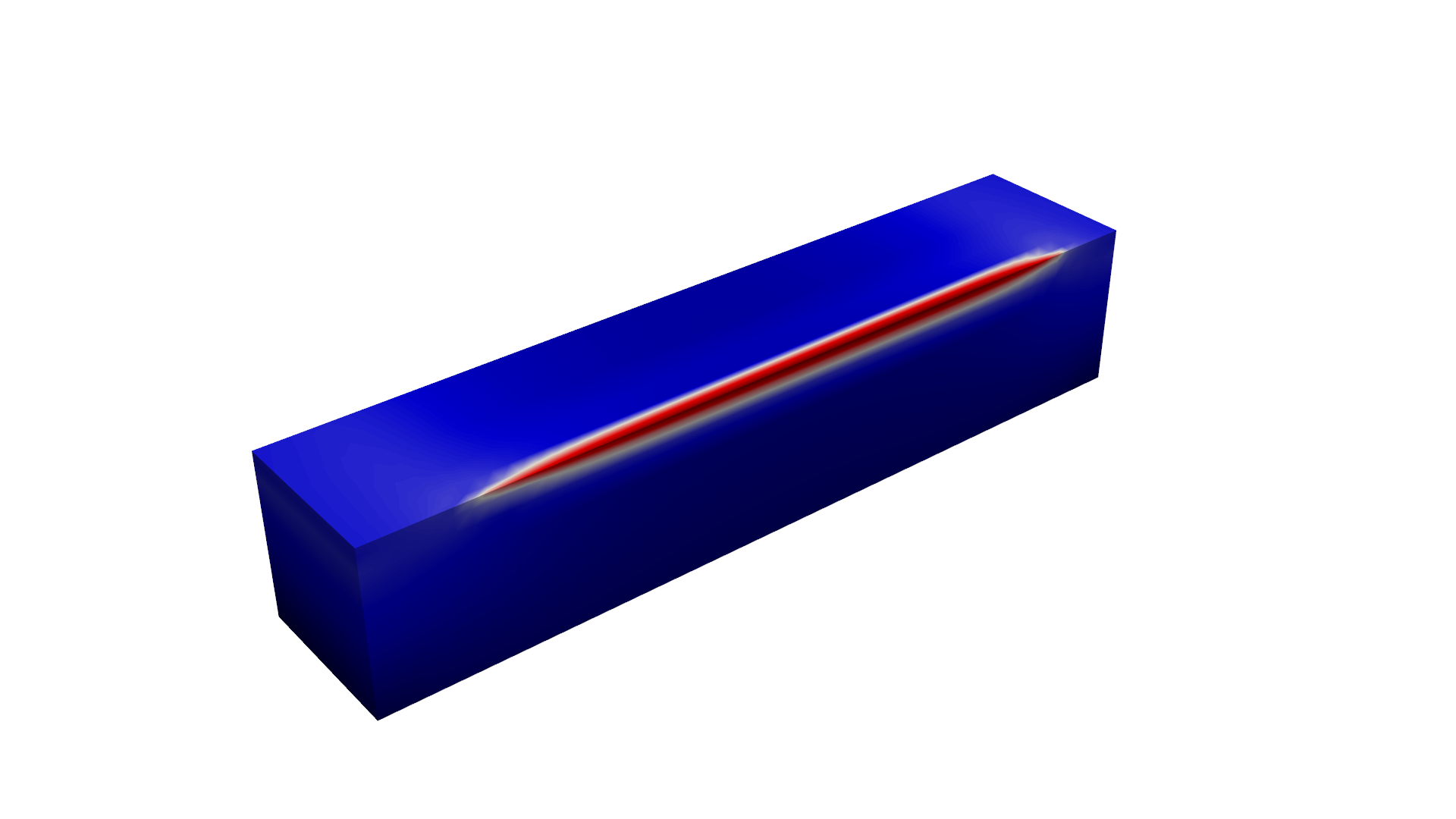}
        \includegraphics[width=0.15\textwidth,trim=0 0 0 0, clip]{graphics/leg_fabian_t20_wave_MPa_pk112}
        \caption{Stress (MPa) prediction $t=\SI{20}{s}$.}\label{fig:after3sf}
    \end{subfigure}
    \caption{Comparison of simulation (target) and surrogate prediction of temperature and stress fields.}
    \label{fig:after3s}
\end{figure}

\subsection{Surrogate model}

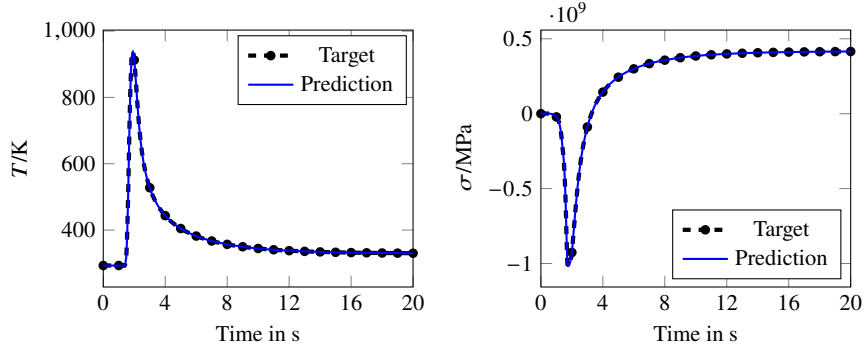
\begin{figure}[tb]
\centering
%\begin{subfigure}[t]{0.49\columnwidth}
%    \centering
\pgfplotsset{
    target/.style={
        dashed, 
        mark=*, mark repeat=10, mark options={solid, fill=black, draw=black, scale=0.5},
        ultra thick, smooth
    },
    pred/.style={
        solid, blue, 
        thick, smooth
    },
}
\begin{tikzpicture}

    \pgfplotstableread[col sep = space]{graphics/data/prediction_at_point.txt}\datatable

    \begin{axis}[
        set layers, mark layer=axis tick labels,    % Draw over markers
        width=5.7cm,
        height=5cm,
        xlabel=Time in \si{s},
        xtick={0,4,8,12,16,20},
        ylabel=$T$/\si{K},
        % ymin=1e-4,
        % ymax=1e-2,
        enlarge x limits=0.,
    ]

        \addplot[target] table[x=ts, y=temp_target] {\datatable};
        \addlegendentry{Target}

        \addplot[pred] table[x=ts, y=temp_pred_n_dim_latent_4] {\datatable};
        \addlegendentry{Prediction}
        
        % \foreach \n in {1,2,3,...,10} {
        %     \addplot[pred] table[x=ts, y=temp_pred_n_dim_latent_\n] {\datatable};
        %     % \addlegendentryexpanded{n=\n}
        % }

    \end{axis}

\end{tikzpicture}
%    \caption{$T$}
%\end{subfigure}
%\begin{subfigure}[t]{0.49\columnwidth}
%    \centering
\pgfplotsset{
    target/.style={
        dashed, 
        mark=*, mark repeat=10, mark options={solid, fill=black, draw=black, scale=0.5},
        ultra thick, smooth
    },
    pred/.style={
        solid, blue, 
        thick, smooth
    },
}
\begin{tikzpicture}

    \pgfplotstableread[col sep = space]{graphics/data/prediction_at_point.txt}\datatable

    \begin{axis}[
        set layers, mark layer=axis tick labels,    % Draw over markers
        width=5.7cm,
        height=5cm,
        xlabel=Time in \si{s},
        xtick={0,4,8,12,16,20},
        ylabel=$\sigma$/\si{MPa},
        % ymin=1e-4,
        % ymax=1e-2,
        enlarge x limits=0.,
        legend pos=south east,
    ]

        \addplot[target] table[x=ts, y=pk11_target] {\datatable};
        \addlegendentry{Target}

        \addplot[pred] table[x=ts, y=pk11_pred_n_dim_latent_4] {\datatable};
        \addlegendentry{Prediction}

    \end{axis}

\end{tikzpicture}
%    \caption{$\sigma$}
%\end{subfigure}
\caption{Temperature and stress predictions at the center point of the DED track.} 
\label{fig:TPKpoint}
\end{figure}

After training the surrogate model as described in \autoref{sec:nodes} and \autoref{sec:aes} with the result data from the thermo-mechanical simulation model, the initial temperature distribution $T(x,y,z,t_0)=T_\text{amb}$ is handed to the surrogate tool center point (TCP) position. \autoref{fig:after3sb} depicts the performance of the temperature surrogate after \SI{3}{s}. Visually, no difference between the ROM and the simulation can be distinguished. 
Next, the derived temperature field is used as an input together with the field data of the 11-stress component $\sigma$ to train the ROM for RS. 
\autoref{fig:after3sd} shows the performance of the stress ROM after \SI{3}{s}, and hardly any difference between the ROM and the simulation can be distinguished. From \SI{4}{s} onwards, when the laser has passed over the block, the cool-down phase starts, and the formed tensile RS become visible more easily, see \autoref{fig:after3sf}.
Similarly, also the evaluation of temperatures and stresses at the center point of the DED track $(0.025,0,0)$, which is shown in \autoref{fig:TPKpoint}, does not exhibit any visual differences between target and surrogate predictions.

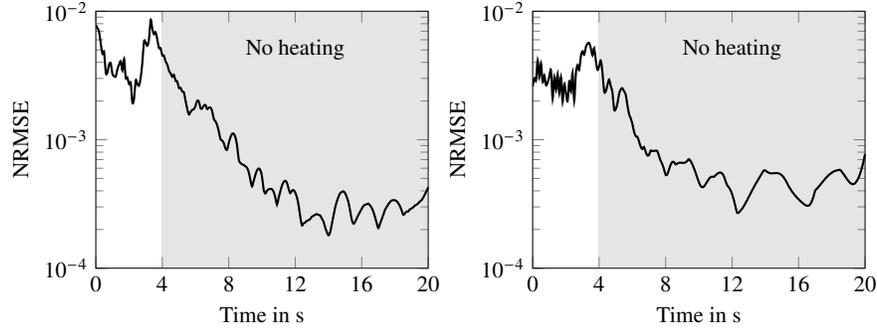
\begin{figure}[tb]
    \centering
    \begin{subfigure}[t]{0.49\columnwidth}
        \centering
        \pgfplotsset{
    NRMSE/.style={
        thick, smooth
    },
}

\tikzsetnextfilename{pk_error_over_time}
\begin{tikzpicture}

    \pgfplotstableread[col sep = space]{graphics/data/nrmses_over_time.txt}\datatable

    \begin{semilogyaxis}[
        % colorbar, 
        % colorbar style={ 
        %     title=Latent dimension, 
        %     point meta min=1, point meta max=10,
        %     ytick={1,...,10}, 
        %     yticklabels={1,2,3,4,5,6,7,8,9,10} 
        %     }, 
        % colormap/jet, 
        % cycle list={[of colormap]},
        width=6cm,
        height=5cm,
        xlabel=Time in \si{s},
        xtick={0,4,8,12,16,20},
        ylabel=NRMSE,
        ymin=1e-4,
        ymax=1e-2,
        enlarge x limits=0.,
    ]
    
        \addplot [forget plot, draw=none, fill=gray, opacity=0.2] coordinates {
            (4 , 1e-4)
            (20, 1e-4)
            (20, 1e-2)
            (4 , 1e-2)
        };
        \node[rectangle, draw=none, rounded corners=5] at (axis cs: 12, 5e-3) {No heating};

        % \foreach \n in {1,2,3,...,10} {
        %     \addplot table[x=ts, y=source_to_pk11_nrmse_n_dim_latent_\n] {\datatable};
        %     % \addlegendentryexpanded{n=\n}
        % }

        \addplot[NRMSE] table[x=ts, y=source_to_temp_nrmse_n_dim_latent_4] {\datatable};

    \end{semilogyaxis}

\end{tikzpicture}
        \caption{Relative prediction error of temperature $T$}\label{fig:errorT}
    \end{subfigure}
    \begin{subfigure}[t]{0.49\columnwidth}
        \centering
        \pgfplotsset{
    NRMSE/.style={
        thick, smooth
    },
}

\tikzsetnextfilename{pk_error_over_time}
\begin{tikzpicture}

    \pgfplotstableread[col sep = space]{graphics/data/nrmses_over_time.txt}\datatable

    \begin{semilogyaxis}[
        % colorbar, 
        % colorbar style={ 
        %     title=Latent dimension, 
        %     point meta min=1, point meta max=10,
        %     ytick={1,...,10}, 
        %     yticklabels={1,2,3,4,5,6,7,8,9,10} 
        %     }, 
        % colormap/jet, 
        % cycle list={[of colormap]},
        width=6cm,
        height=5cm,
        xlabel=Time in \si{s},
        xtick={0,4,8,12,16,20},
        ylabel=NRMSE,
        ymin=1e-4,
        ymax=1e-2,
        enlarge x limits=0.,
    ]
    
        \addplot [forget plot, draw=none, fill=gray, opacity=0.2] coordinates {
            (4 , 1e-4)
            (20, 1e-4)
            (20, 1e-2)
            (4 , 1e-2)
        };
        \node[rectangle, draw=none, rounded corners=5] at (axis cs: 12, 5e-3) {No heating};

        % \foreach \n in {1,2,3,...,10} {
        %     \addplot table[x=ts, y=source_to_pk11_nrmse_n_dim_latent_\n] {\datatable};
        %     % \addlegendentryexpanded{n=\n}
        % }

        \addplot[NRMSE] table[x=ts, y=source_to_pk11_nrmse_n_dim_latent_4] {\datatable};

    \end{semilogyaxis}

\end{tikzpicture}
        \caption{Relative prediction error of stress $\sigma$} \label{fig:errorpk1}
    \end{subfigure}
    \caption{Mean absolute prediction errors over time for all field points. Note that there is no heating after $t=\SI{4}{s}$.}
    \label{fig:errors}
\end{figure}

To demonstrate the accuracy of the models in more detail, \autoref{fig:errors} shows the relative errors over time for the temperature and stress fields, which are both consistently less than \SI{1}{\%}.
Here, the root mean squared error normalized with the standard deviation of the full trajectory is used. For a prediction $\Tilde{y}_{ij}=\Tilde{y}(t_i,x_j)$ and corresponding ground truth $y_{ij}=y(t_i,x_j)$ the $\NRMSE$ for time $t_k$ is calculated via the following formula:
\begin{equation}
    \NRMSE(t_k) = \frac{\RMSE(t_k)}{\STD} = \sqrt{\frac{\sum_j^{n_s}(\Tilde{y}_{kj} - y_{kj})^2}{\frac{1}{n_t} \sum_i^{n_t}\sum_j^{n_s}(\Bar{y} - y_{ij})^2}}
\end{equation}
Here, $\Bar{y}$ denotes the mean of $y$.
For the total error of an entire trajectory prediction the $\RMSE$ is computed with the mean over all time and space samples:
\begin{equation}
    \NRMSE = \frac{\RMSE}{\STD} = \sqrt{\frac{\sum_i^{n_t}\sum_j^{n_s}(\Tilde{y}_{ij} - y_{ij})^2}{\sum_i^{n_t}\sum_j^{n_s}(\Bar{y} - y_{ij})^2}}
\end{equation}

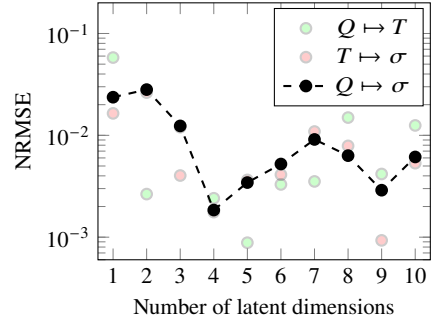
\begin{figure}[tb]
    \centering
    \pgfplotsset{
    main/.style={
        thick,
        dashed,
        mark=*, mark options={solid, draw=black, fill=black, scale=1}
    },
    side/.style={
        main,
        only marks,
        opacity=0.2,
    },
}

\tikzsetnextfilename{pk_error_over_time}
\begin{tikzpicture}

    \pgfplotstableread[col sep = space]{graphics/data/total_banrmses.txt}\datatable

    \begin{semilogyaxis}[
        width=6cm,
        height=5cm,
        xlabel=Number of latent dimensions,
        xtick={1,2,...,10},
        ylabel=NRMSE,
        ymin=6e-4,
        ymax=2e-1,
        enlarge x limits=0.05,
    ]

        \addplot[side, mark options={solid, fill=green}] table[x=n_dim_latent, y=source_to_temp_banrmse] {\datatable};
        \addlegendentry{$Q\mapsto T$}
        \addplot[side, mark options={solid, fill=red}] table[x=n_dim_latent, y=temp_to_pk11_banrmse] {\datatable};
        \addlegendentry{$T\mapsto\sigma$}
        \addplot[main] table[x=n_dim_latent, y=source_to_pk11_banrmse] {\datatable};
        \addlegendentry{$Q\mapsto\sigma$}
    
    \end{semilogyaxis}

\end{tikzpicture}
    \caption{Errors per number of latent dimensions for the case of predicting temperatures from the heat source (green), stresses from temperatures (red) and the full evaluation chain for stresses from heat sources (black).}
    \label{fig:latentstudy}
\end{figure}

Furthermore, a study of a variable number of latent degrees of freedom, see \autoref{fig:latentstudy}, reveals that employing $n_l=4$ degrees of freedom provides the best overall performance. 
Consequently, the evolution of the temperature field can be replicated very accurately and quickly in the latent space NODE.
Thus, this model was used for all results shown above.
The required inference time, including projection onto the full space, was $\SI{412}{ms} \pm \SI{13.6}{ms}$ on a standard desktop CPU for a prediction of \SI{20}{s} real-time. 
The prediction of the Cauchy stress component required $\SI{410}{ms} \pm \SI{9.92}{ms}$. 

In summary, the combination of autoencoders and NODEs captured the advective behavior with very satisfactory accuracy using only four latent variables.
Moreover, all results have also been visualized in a video\footnote{\url{http://dt4ded.maxkann.de}}.

\section{Summary and outlook}
\label{sec:summary}

This work addresses the growing demand for digital twin frameworks in AM, focusing on enhancing real-time prediction capabilities for complex phenomena such as temperature and RS evolution in DED processes. A comprehensive review of the current state of the art reveals a significant gap in integrating ML models within DT frameworks, particularly for the accurate and rapid prediction of RS. While previous approaches have advanced in some respects, none have utilized NODEs for AM predictive modeling.

To bridge this gap, this study presents a pioneering proof of concept, leveraging NODEs to replicate temperature profiles and RS distributions in a single-line DED deposition scenario. The model demonstrates excellent precision, achieving temperature predictions with less than 1\% relative error in under 0.5 seconds. Moreover, RS prediction also achieves a relative error of less than 1\% within the critical time window of stress formation. These results underscore the potential of NODEs to enhance predictive accuracy and efficiency, advancing the feasibility of real-time digital twin applications in AM.

%\subsection{Outlook}
The promising outcomes of this study suggest the potential application of the proposed NODE-based concept within a digital twin framework, enabling live control of RS during additive manufacturing. However, achieving this ambitious goal requires addressing several future challenges.

To align the model with real-world experiments, it is essential to incorporate process parameters such as laser power, wire feed rate, two-dimensional TCP paths, and the distance to the built part. These inputs will further necessitate the inclusion of derived parameters, such as deposition width and height, which must be integrated into a material deposition approach, for instance, using the quiet element method~\cite{nain_dedthesis_2022}. 
The thermo-mechanical problem necessitates a total Lagrangian formulation for scenarios involving thin substrates or parts with high geometric complexity of large deformations. The nonlinear material model might benefit from advanced machine learning modeling that includes physical objectivity priors~\cite{Klein2023,zlati23}. 

A significant challenge lies in extending the material deposition model to account for thermal interactions with the surrounding environment accurately. Prior research has highlighted the critical role of convective and radiative heat transfer effects, which must be modeled in a conjugate fashion~\cite{kannapinn_digital_2023, kannapinn_twinlab_2024,kannapinn_wine_2024}. Furthermore, the influence of forced protective gas flow on localized heat transfer demands dedicated investigation. In additive manufacturing, an additional complexity arises from the sequential nature of layer formation. Newly deposited layers exchange heat (via convection and radiation) with the environment until subsequent layers are added. Heat conduction dominates in the contact zones between layers. At the same time, air properties must be modeled in unprinted areas and material-specific properties, such as those of Inconel 718, must be incorporated only after deposition.

In surrogate modeling with NODEs, the next logical steps include rigorous interpolation and extrapolation tests of the framework under varying process parameters, such as deposition speed and laser power. To ensure data efficiency, developing optimized designs of experiments will be critical~\cite{kannapinn_twinlab_2024}.

\begin{acknowledgement}
The authors acknowledge the support of the Graduate School CE within the Centre for Computational Engineering at the Technical University of Darmstadt. 
\end{acknowledgement}
\ethics{Competing Interests}{
The authors have no conflicts of interest to declare that are relevant to the content of this chapter.}

\renewcommand*{\bibfont}{\footnotesize}
\printbibliography

\clearpage

\end{document}